\documentclass{PoS}
\usepackage{subfig}
\usepackage{float}
\usepackage{url}
\usepackage{braket}
\newcommand{\Tr}{\mbox{\rm Tr}}
\newcommand{\ReC}{\mbox{\rm Re}}

\title{Chromoelectric and chromomagnetic fields for the static gluon-quark-antiquark system}

\ShortTitle{Chromoelectric and chromomagnetic fields for the static gluon-quark-antiquark system}

\author{Marco Cardoso \\
CFTP, Instituto Superior Técnico\\
E-mail: \email{mjdcc@cftp.ist.utl.pt}
}

\author{\speaker{Nuno Cardoso} \\
CFTP, Instituto Superior Técnico\\
E-mail: \email{nunocardoso@cftp.ist.utl.pt}
}

\author{Pedro Bicudo \\
CFTP, Instituto Superior Técnico\\
E-mail: \email{bicudo@ist.utl.pt}
}

\abstract{
The chromoelectric and chromomagnetic fields, created by a static gluon-quark-antiquark system, are computed in the quenched approximation of lattice QCD, in a $24^3\times 48$ lattice at $\beta=6.2$. We study two geometries, one with a U shape and another with an L shape. The degenerate case of the two gluon glueball is also studied. This is relevant to understand the microscopic structure of hadrons, in particular of hybrids. This also contributes to understand confinement with flux tubes of the chromoelectric field, and to discriminate between the models of fundamental or adjoint tubes.
}

\FullConference{The XXVII International Symposium on Lattice Field Theory - LAT2009\\
		 July 26-31 2009\\
		 Peking University, Beijing, China}

\begin{document}

\section{Introduction}

\begin{figure}[H]
\begin{centering}
    \subfloat[U shape geometry.\label{fig:shapeU}]{
\begin{centering}
    \includegraphics[height=4cm]{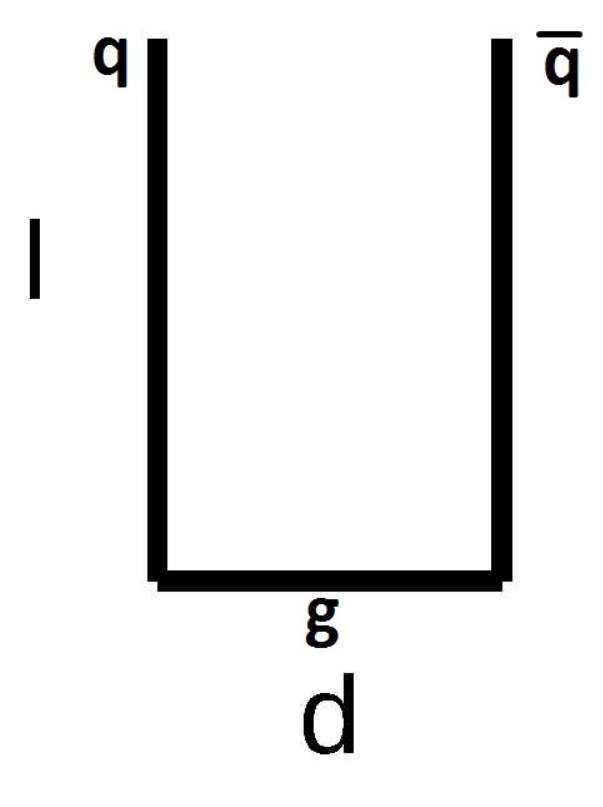}
\par\end{centering}}
    \subfloat[L shape geometry.\label{fig:shapeL}]{
\begin{centering}
    \includegraphics[height=4cm]{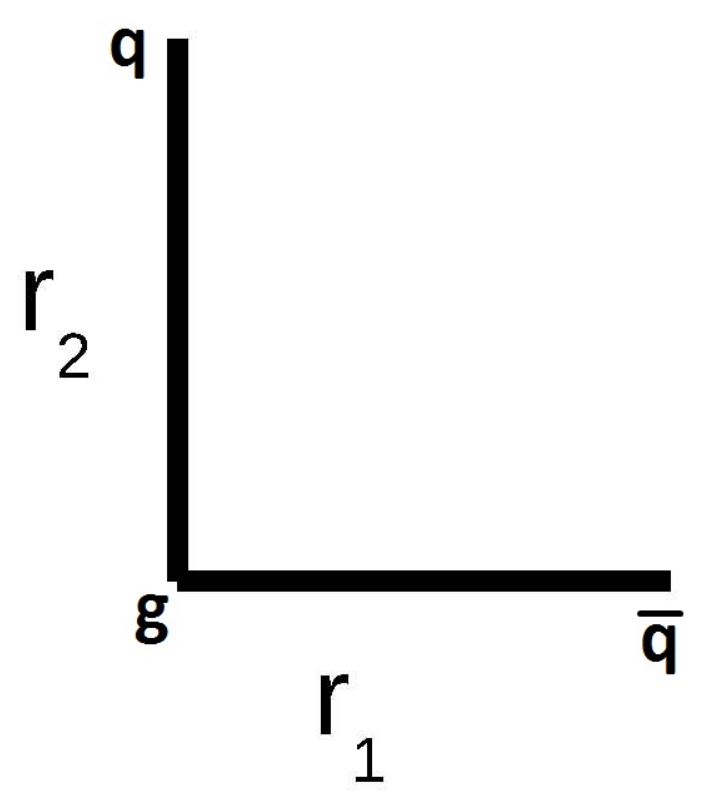}
\par\end{centering}}
\par\end{centering}
    \caption{gluon-quark-antiquark geometries, U and L shapes.}
    \label{shape}
\end{figure}

The aim of this work is to compute the chromoelectric and chromomagnetic fields and study the shape of the flux-tubes for the hybrid gluon-quark-antiquark system.
We study two different geometries for the hybrid system, one with a U shape and another with L shape, figure. \ref{shape}.

\section{The Wilson Loops and Chromo Fields for exotic systems}
\begin{figure}[h]
\begin{centering}
    \subfloat[\label{loop0}]{
\begin{centering}
    \includegraphics[width=5cm]{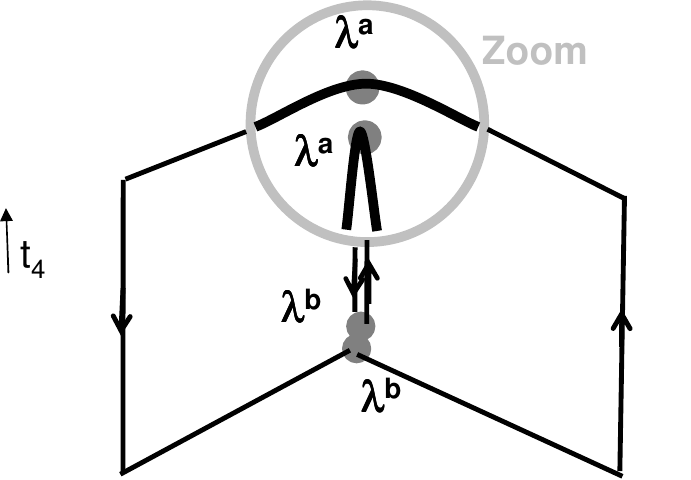}
\par\end{centering}}
    \subfloat[\label{loop1}]{
\begin{centering}
    \includegraphics[width=7cm]{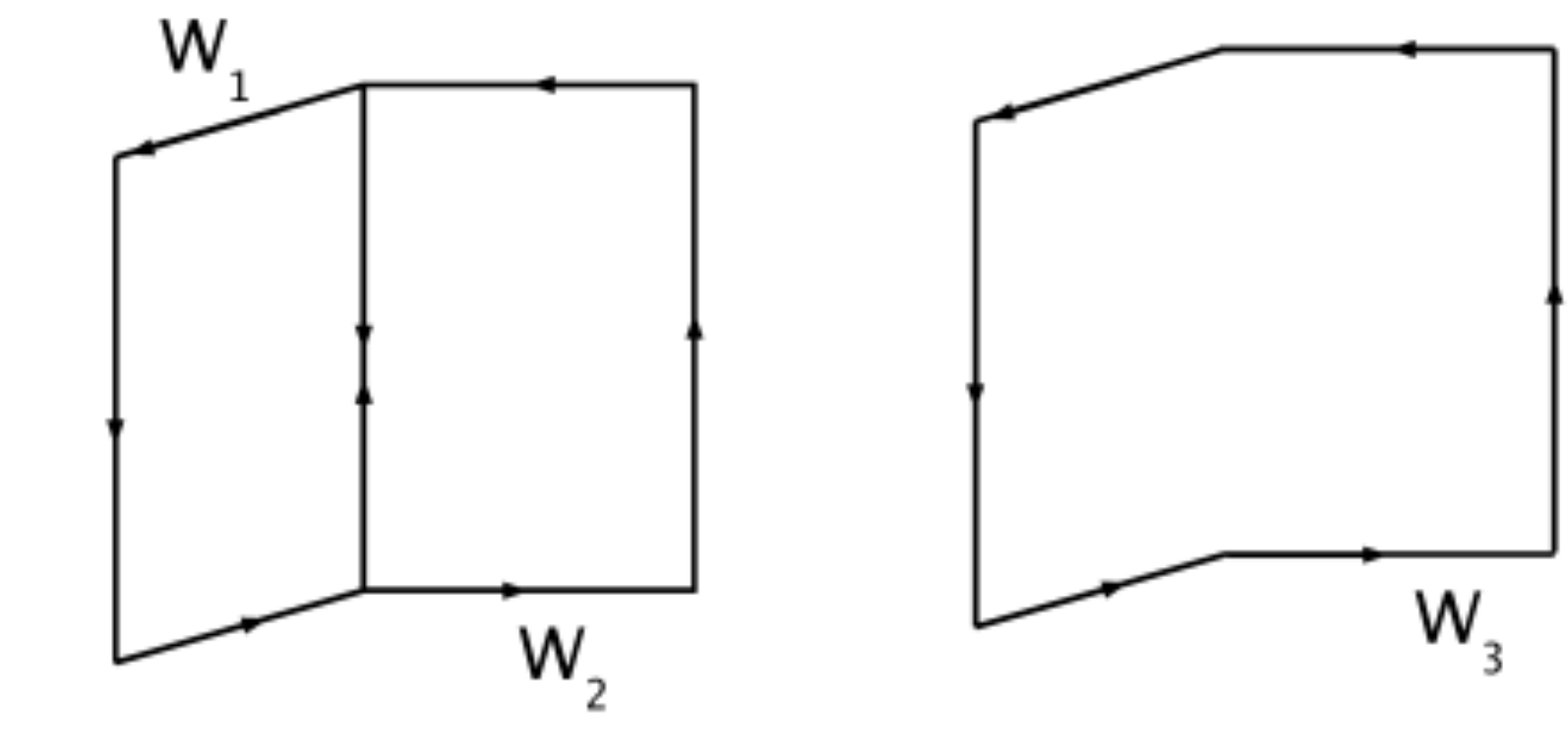}
\par\end{centering}}
\par\end{centering}
    \caption{\subref{loop0} Wilson loop for the $gq\overline{q}$ and equivalent position of the static antiquark, gluon, and quark. \subref{loop1} Simple Wilson loops that make the $gq\overline{q}$ Wilson loop.}
    \label{loop}
\end{figure}

In principle, any Wilson loop with a geometry similar to that represented in figure \ref{loop0}, describing correctly the quantum numbers of the hybrid, is appropriate, although the signal to noise ratio may depend on the choice of the Wilson loop.
A correct Wilson loop must include an SU(3) octet (the gluon), an SU(3) triplet (the quark) and an SU(3) antitriplet (the antiquark), as well as the connection between the three links of the gluon, the quark and the antiquark.

The Wilson loop for this system is given by:
\begin{eqnarray}
    W_{gq\overline{q}} & = & \frac{1}{16}Tr[\lambda^b T^+ \lambda^a T]Tr[\lambda^b S_1 \lambda^a S_2]\\
    & = & W_1 W_2 - \frac{1}{3}W_3
\end{eqnarray}
where $W_1$, $W_2$ and $W_3$ are the simple Wilson loops shown in figure \ref{loop1}.
When $r_1=0$, $W_1=3$ and $W_2=W_3$, the operator reduces to the mesonic Wilson loop
and when $\mu=\nu$ and $r_1=r_2=r$, $W_2=W_1^{\dagger}$ and $W_3=3$, $W_{gq\overline{q}}$ reduces
to $W_{gq\overline{q}}(r,r,t)=|W(r,t)|^2-1$, that is the Wilson loop in the adjoint representation
used to compute the potential between two static gluons.

In order to improve the signal to noise ratio of the Wilson loop, the links are replaced by "fat links",
\begin{eqnarray}
    U_{\mu}\left(s\right) & \rightarrow & P_{SU(3)}\frac{1}{1+6w}\left(U_{\mu}\left(s\right) +w\sum_{\mu\neq\nu}U_{\nu}\left(s\right)U_{\mu}\left(s+\nu\right)U_{\nu}^{\dagger}\left(s+\mu\right)\right)
\end{eqnarray}

We use $w = 0.2$ and iterate this procedure 25 times in the spatial direction.

We ontain the chromoelectric and chromomagnetic fields on the lattice, by using:
\begin{equation}
    \Braket{E^2}= \Braket{P_{0i}}-\frac{\Braket{W\,P_{0i}}}{\Braket{W}}
\end{equation}
\begin{equation}
    \Braket{B^2}= \frac{\Braket{W\,P_{ij}}}{\Braket{W}}-\Braket{P_{ij}}
\end{equation}
and the plaquette is given by
\begin{equation}
P_{\mu\nu}\left(s\right)=1 - \frac{1}{3} \ReC\Tr\left[ U_{\mu}(s) U_{\nu}(s+\mu) U_{\mu}^\dagger(s+\nu) U_{\nu}^\dagger(s) \right]
\end{equation}
We only apply the smearing technique to the Wilson loop.
The relationship between various plaquettes and the field components is:
\begin{center}
\begin{tabular}{cc}
component & plaquette\tabularnewline
\hline
 $E_x$ & $P_{xt}$ \tabularnewline
 $E_y$ & $P_{yt}$ \tabularnewline
 $E_z$ & $P_{zt}$ \tabularnewline
 $B_x$ & $P_{yz}$ \tabularnewline
 $B_y$ & $P_{zx}$ \tabularnewline
 $B_z$ & $P_{xy}$ \tabularnewline
\end{tabular}
\end{center}

The energy ($\varepsilon$) and action ($\gamma$) density is given by
\begin{equation}
    \varepsilon = \frac{1}{2}\left( \Braket{E^2} + \Braket{B^2}\right)
\end{equation}
\begin{equation}
    \gamma = \frac{1}{2}\left( \Braket{E^2} - \Braket{B^2}\right)
\end{equation}

\section{Results for the gluon-quark-antiquark}
Here we present the results of our simulations with 141 $24^3 \times 48$, $\beta = 6.2$ configurations generated with the version 6 of the MILC code \cite{MILC}, via a combination of Cabbibo-Mariani and overrelaxed updates. The results are presented in lattice spacing units.

\begin{figure}[H]
\begin{centering}
    \subfloat[Chromoelectric Field\label{fig:cfield_U_d_0_l_8_E}]{
\begin{centering}
    \includegraphics[width=3.5cm]{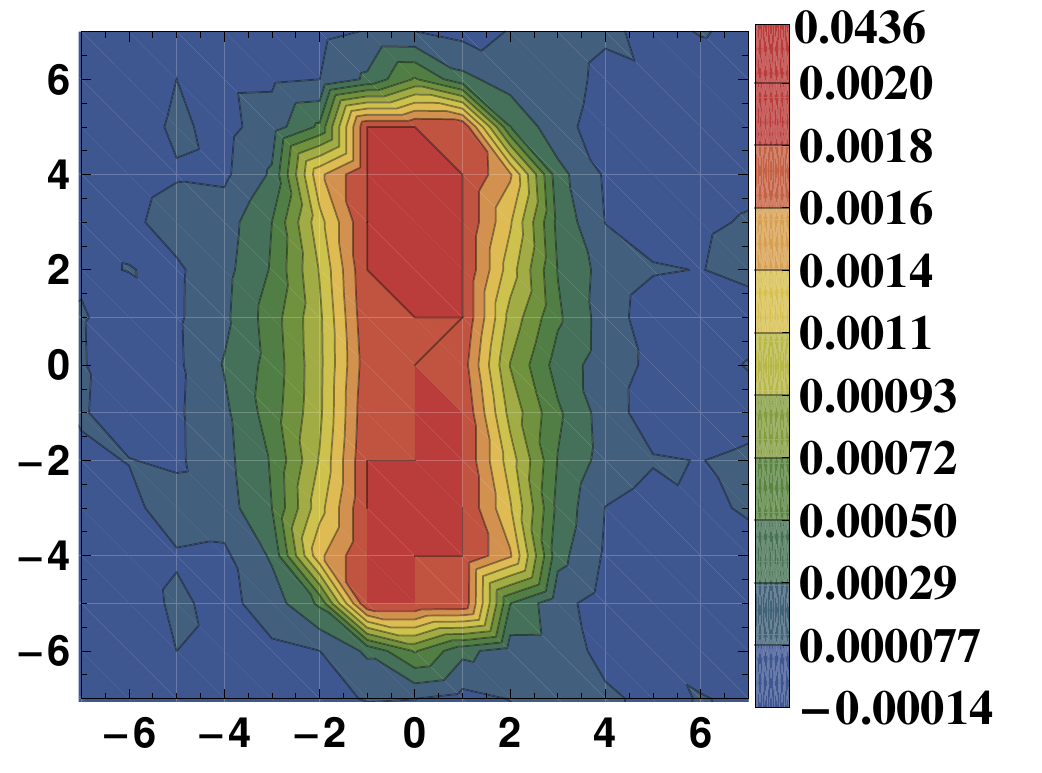}
\par\end{centering}}
    \subfloat[Chromomagnetic Field\label{fig:cfield_U_d_0_l_8_B}]{
\begin{centering}
    \includegraphics[width=3.5cm]{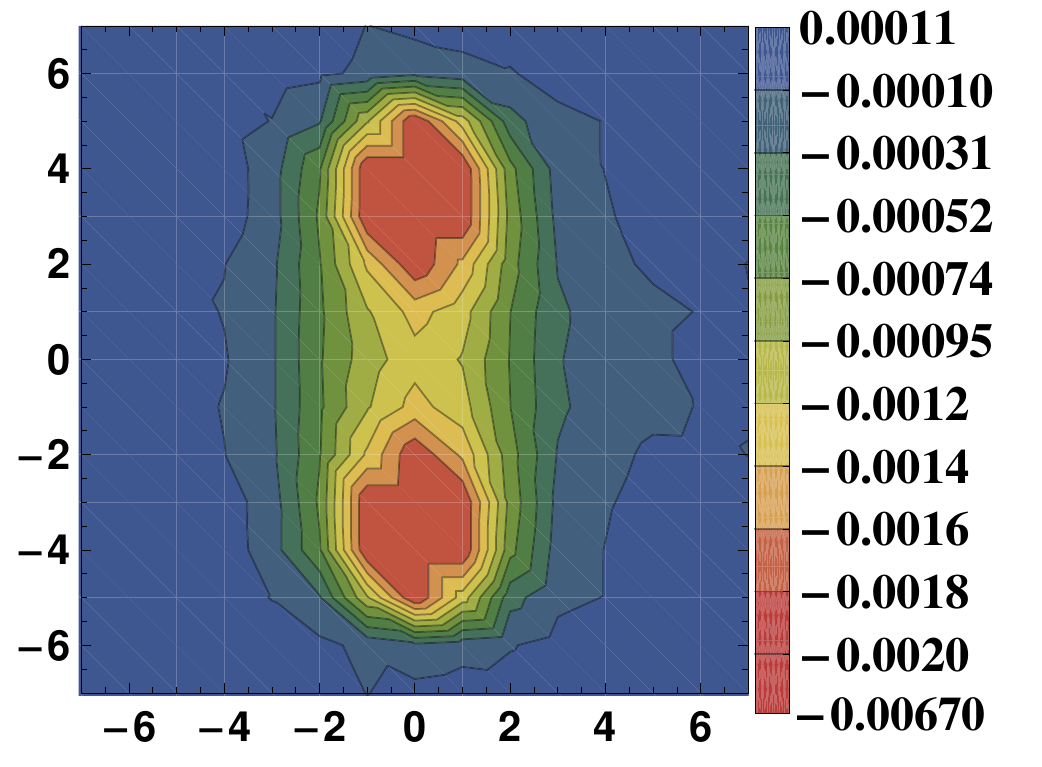}
\par\end{centering}}
    \subfloat[Energy Density\label{fig:cfield_U_d_0_l_8_Energ}]{
\begin{centering}
    \includegraphics[width=3.5cm]{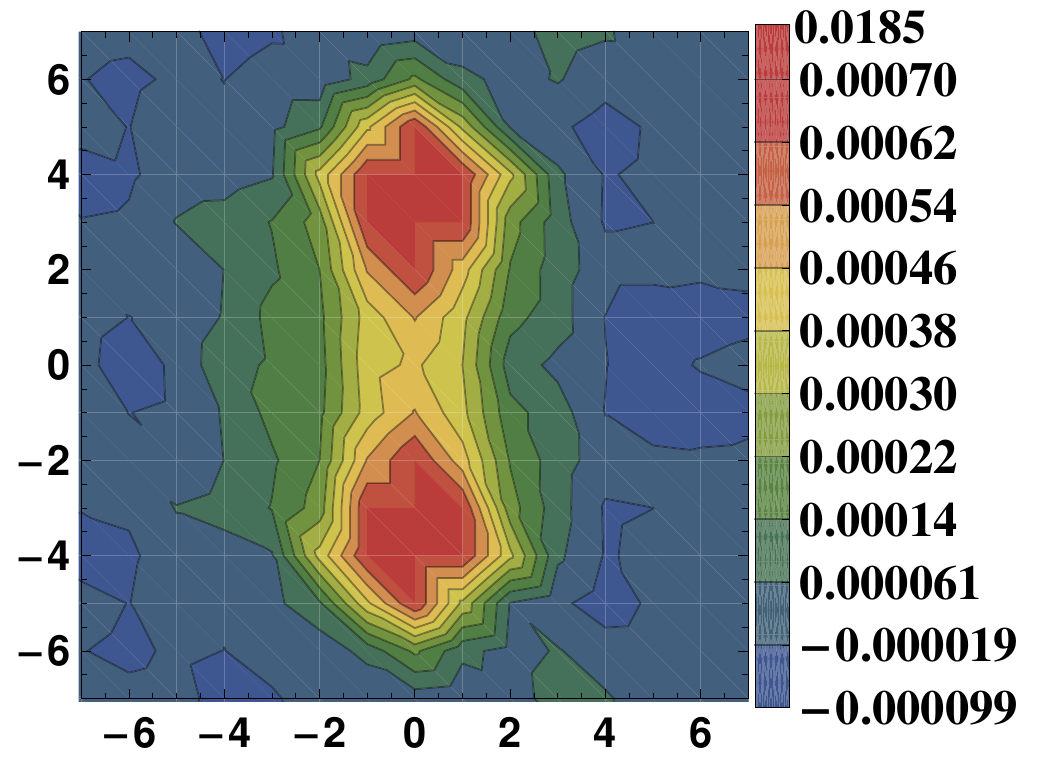}
\par\end{centering}}
    \subfloat[Action Density\label{fig:cfield_U_d_0_l_8_Act}]{
\begin{centering}
    \includegraphics[width=3.5cm]{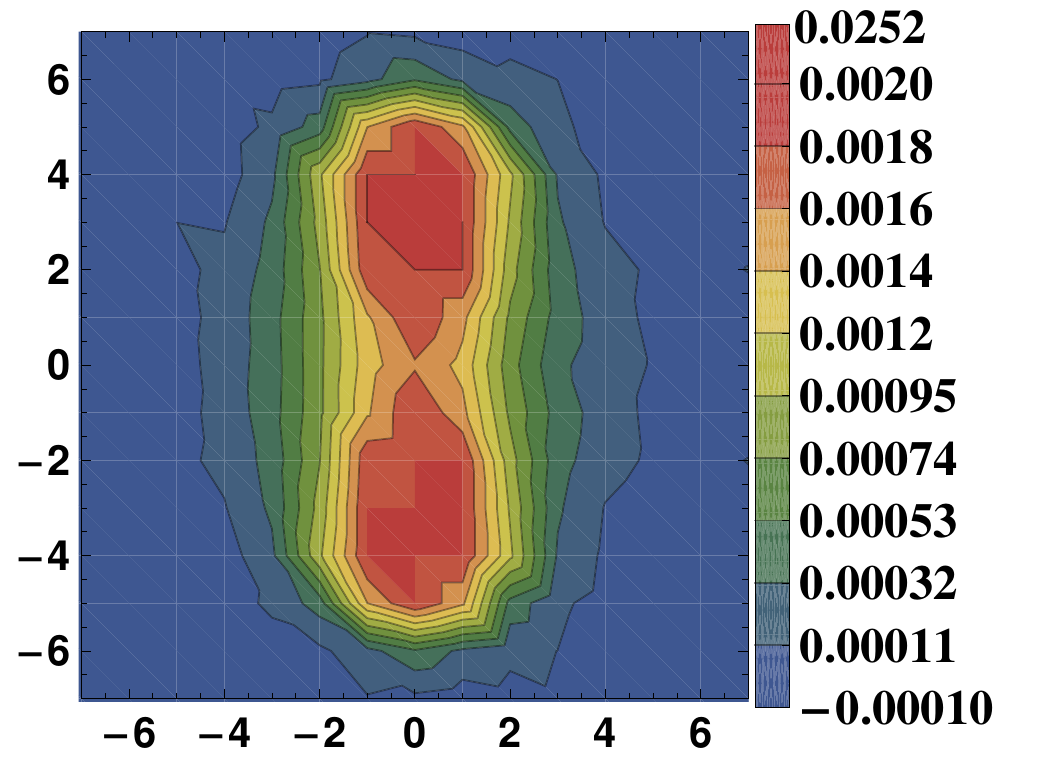}
\par\end{centering}}
\par\end{centering}
    \caption{Results for the U geometry with $d=0$ and $l=8$. In this case, the quark and the antiquark are superposed and the results are consistent
with the degenerate case of the two gluon glueball.}
    \label{cfield_U_d_0_l_8}
\end{figure}

\begin{figure}[H]
\begin{centering}
    \subfloat[Chromoelectric Field\label{fig:cfield_U_d_2_l_8_E}]{
\begin{centering}
    \includegraphics[width=3.5cm]{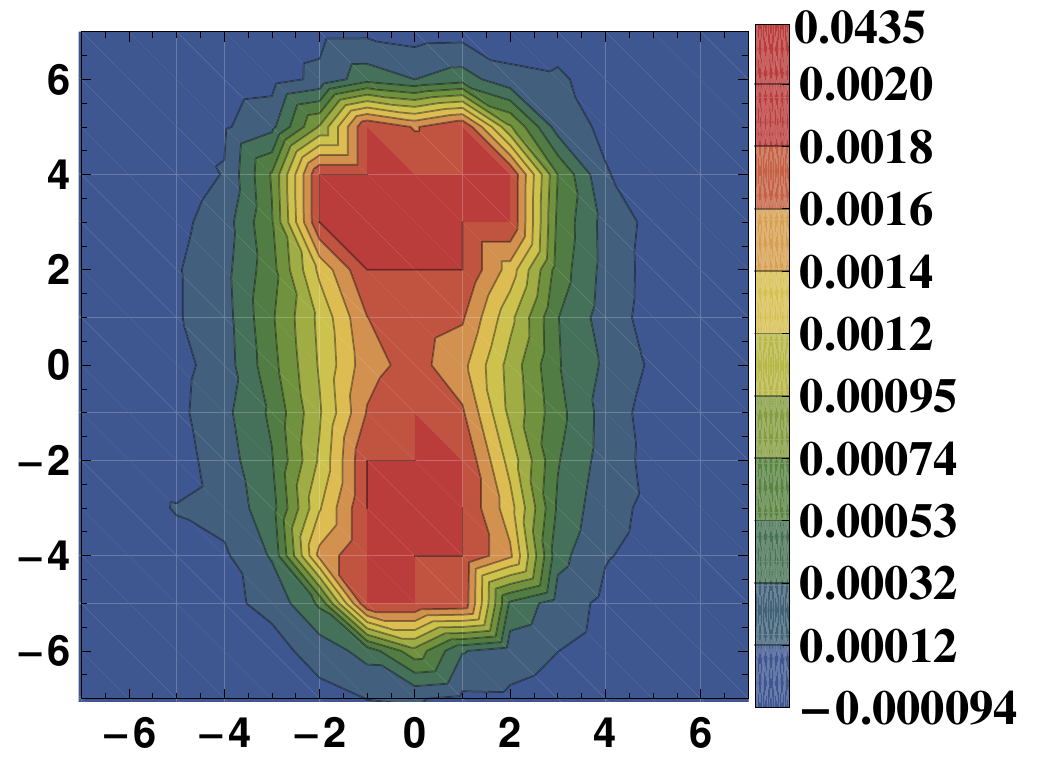}
\par\end{centering}}
    \subfloat[Chromomagnetic Field\label{fig:cfield_U_d_2_l_8_B}]{
\begin{centering}
    \includegraphics[width=3.5cm]{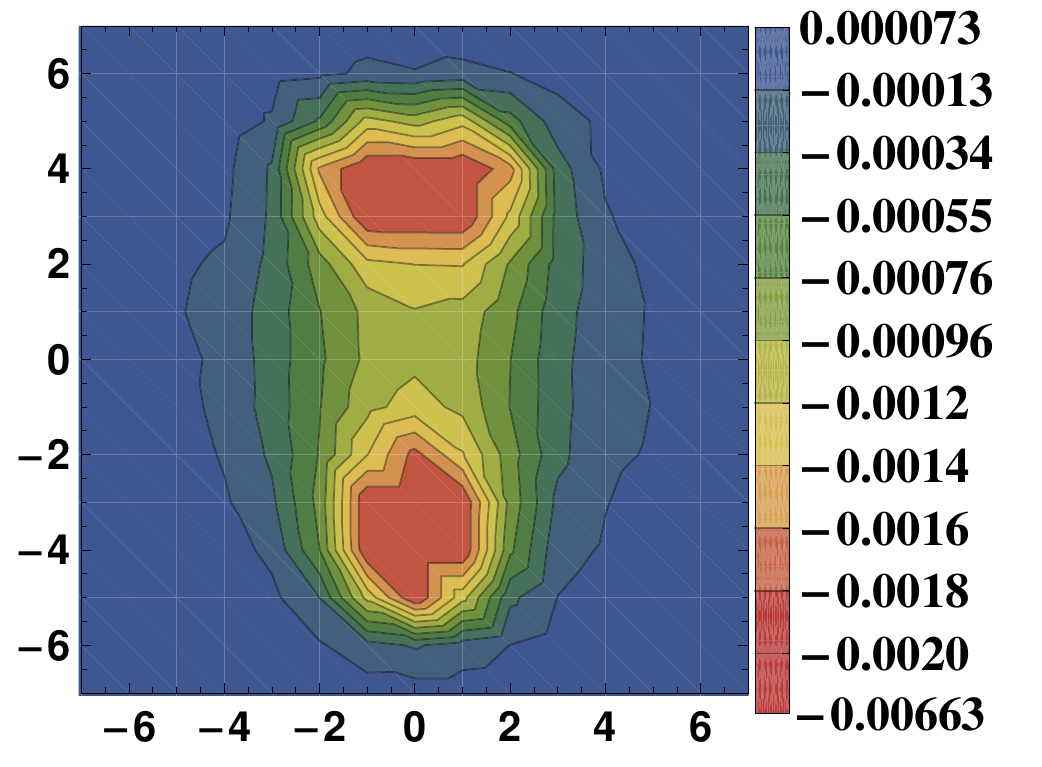}
\par\end{centering}}
    \subfloat[Energy Density\label{fig:cfield_U_d_2_l_8_Energ}]{
\begin{centering}
    \includegraphics[width=3.5cm]{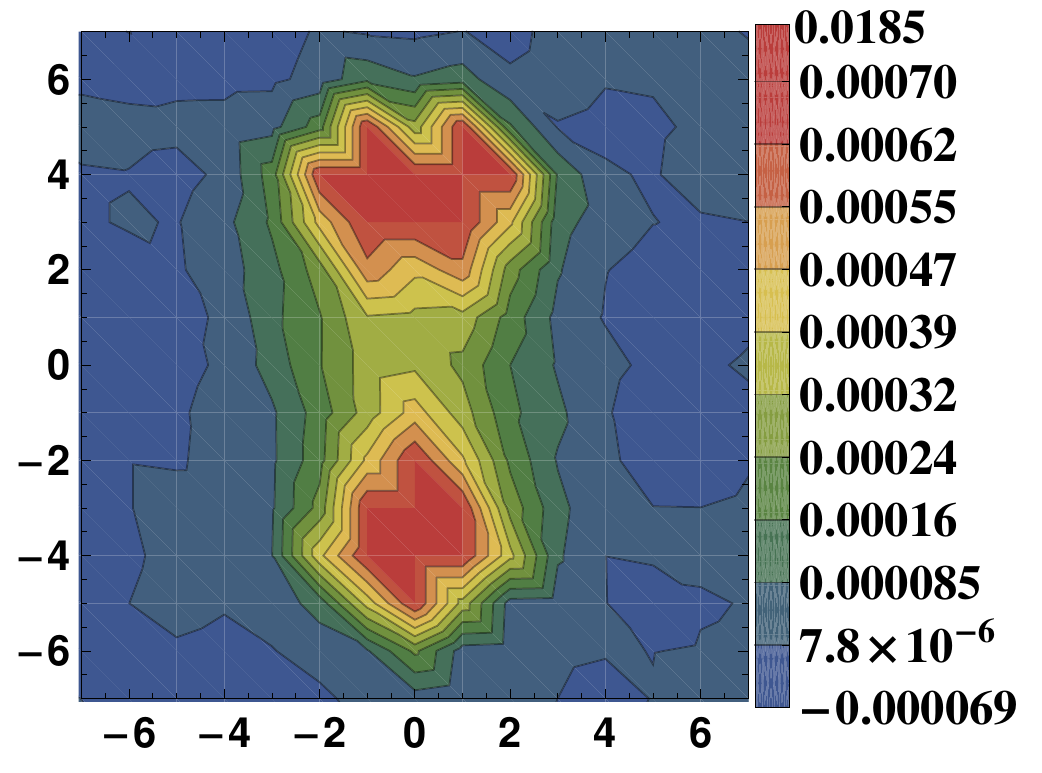}
\par\end{centering}}
    \subfloat[Action Density\label{fig:cfield_U_d_2_l_8_Act}]{
\begin{centering}
    \includegraphics[width=3.5cm]{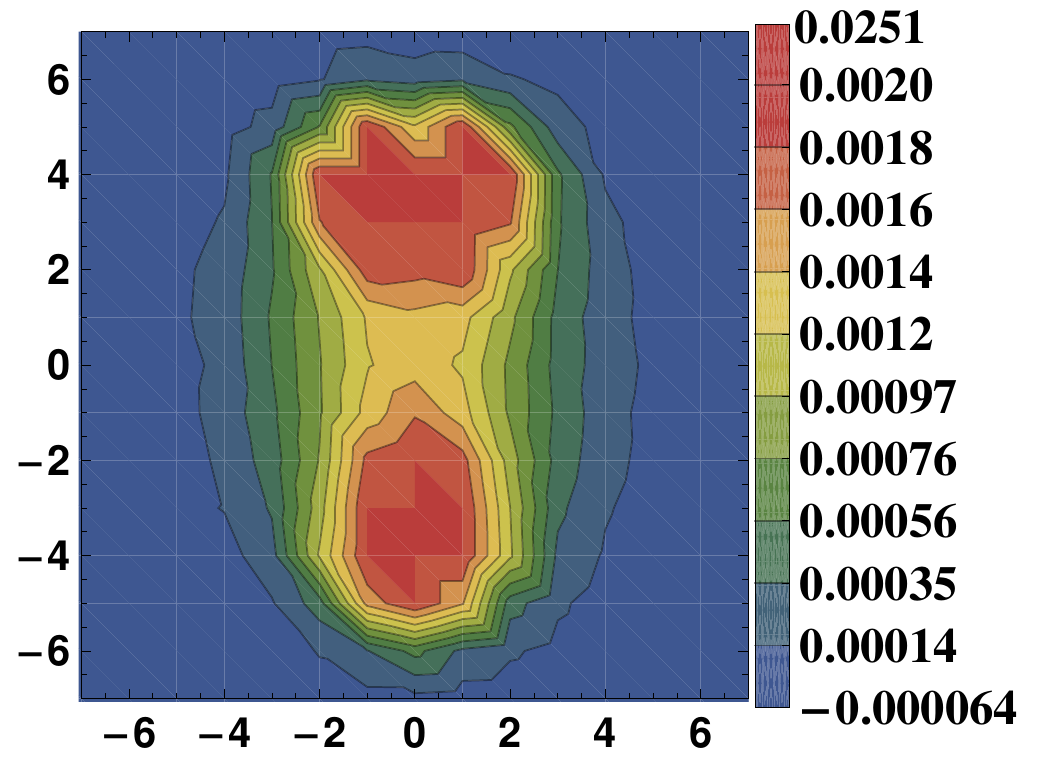}
\par\end{centering}}
\par\end{centering}
    \caption{Results for the U geometry with $d=2$ and $l=8$}
    \label{cfield_U_d_2_l_8}
\end{figure}

\begin{figure}[H]
\begin{centering}
    \subfloat[Chromoelectric Field\label{fig:cfield_U_d_4_l_8_E}]{
\begin{centering}
    \includegraphics[width=3.5cm]{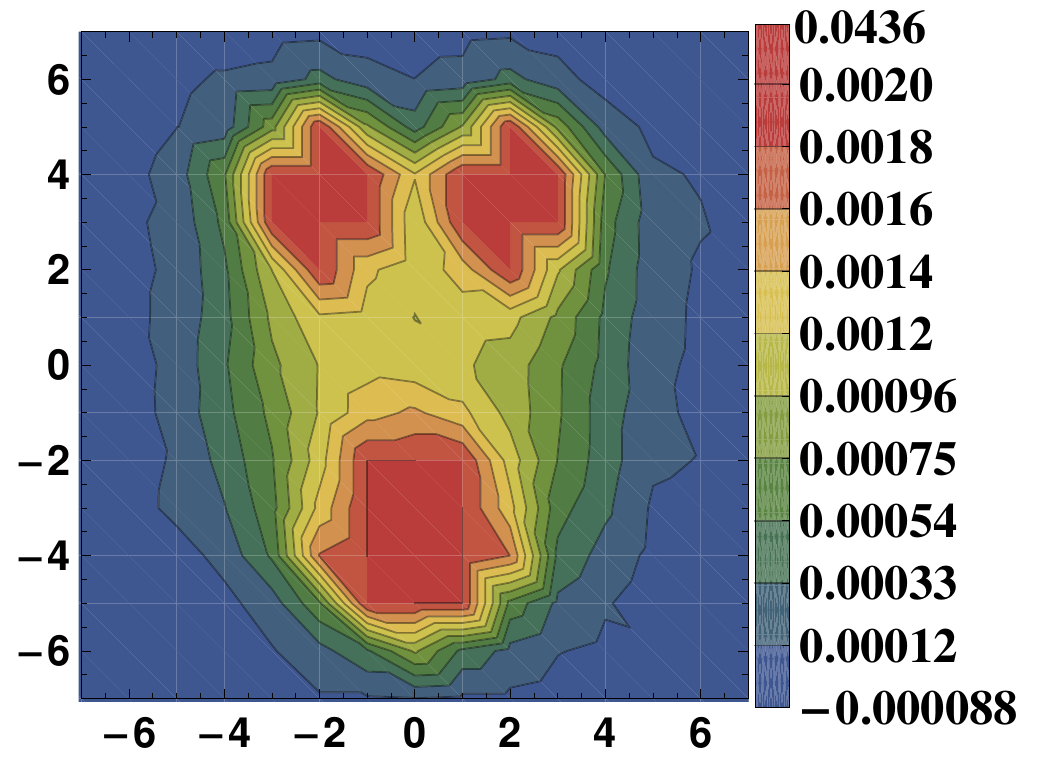}
\par\end{centering}}
    \subfloat[Chromomagnetic Field\label{fig:cfield_U_d_4_l_8_B}]{
\begin{centering}
    \includegraphics[width=3.5cm]{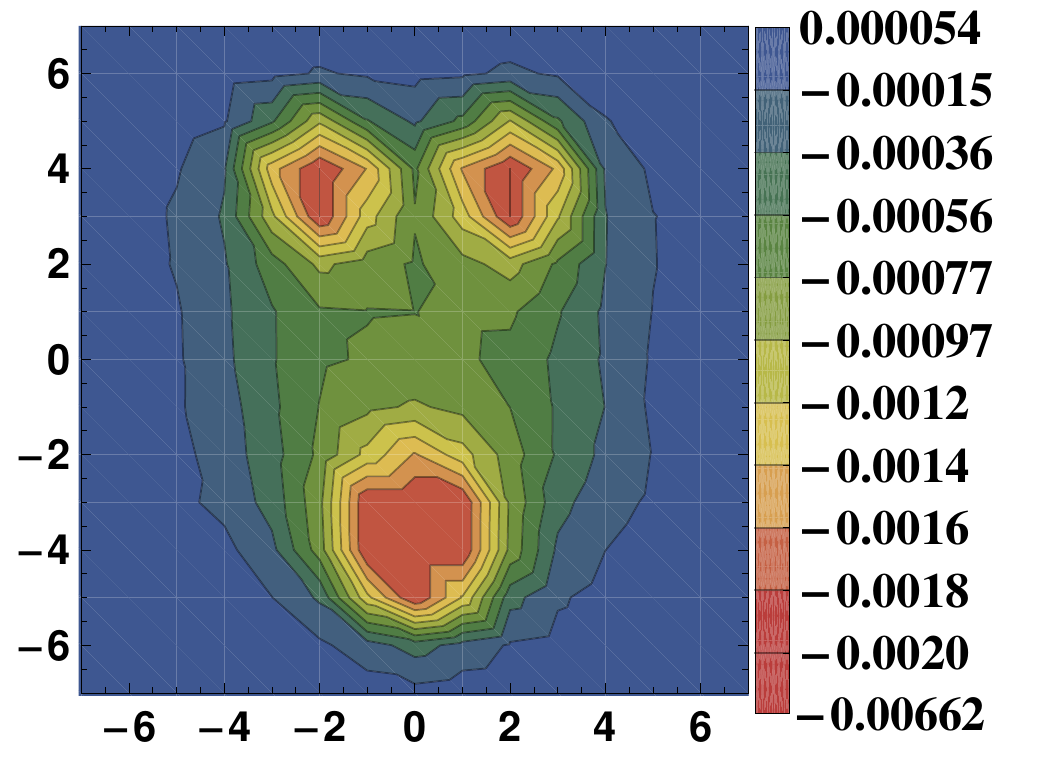}
\par\end{centering}}
    \subfloat[Energy Density\label{fig:cfield_U_d_4_l_8_Energ}]{
\begin{centering}
    \includegraphics[width=3.5cm]{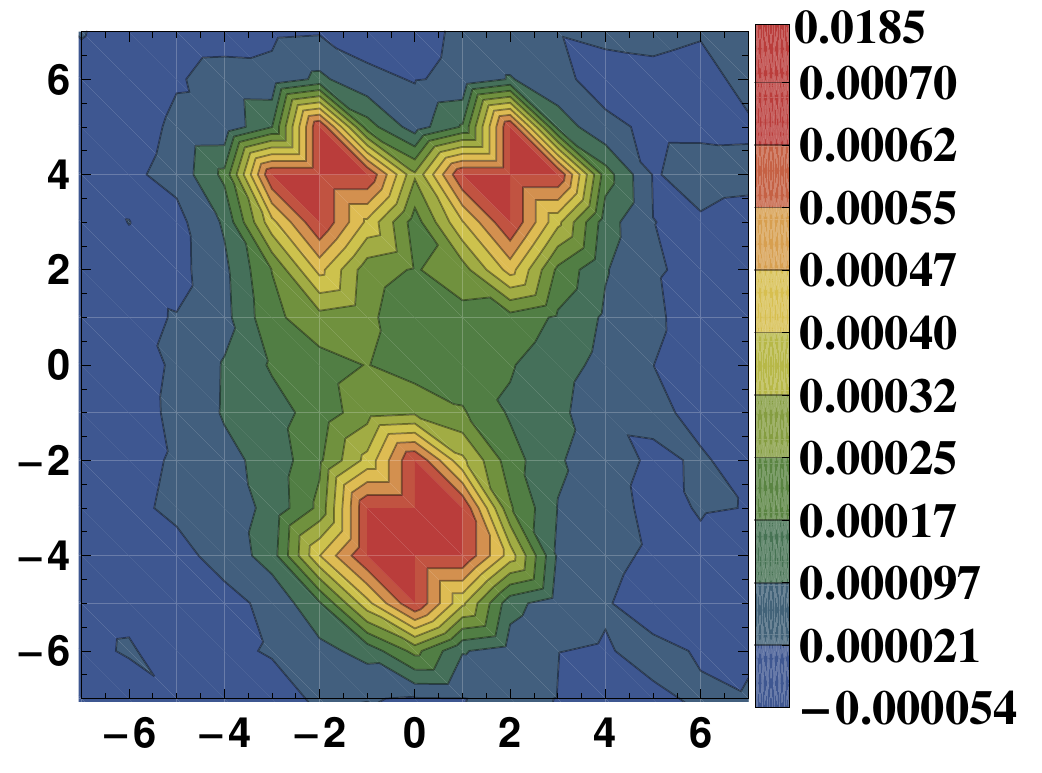}
\par\end{centering}}
    \subfloat[Action Density\label{fig:cfield_U_d_4_l_8_Act}]{
\begin{centering}
    \includegraphics[width=3.5cm]{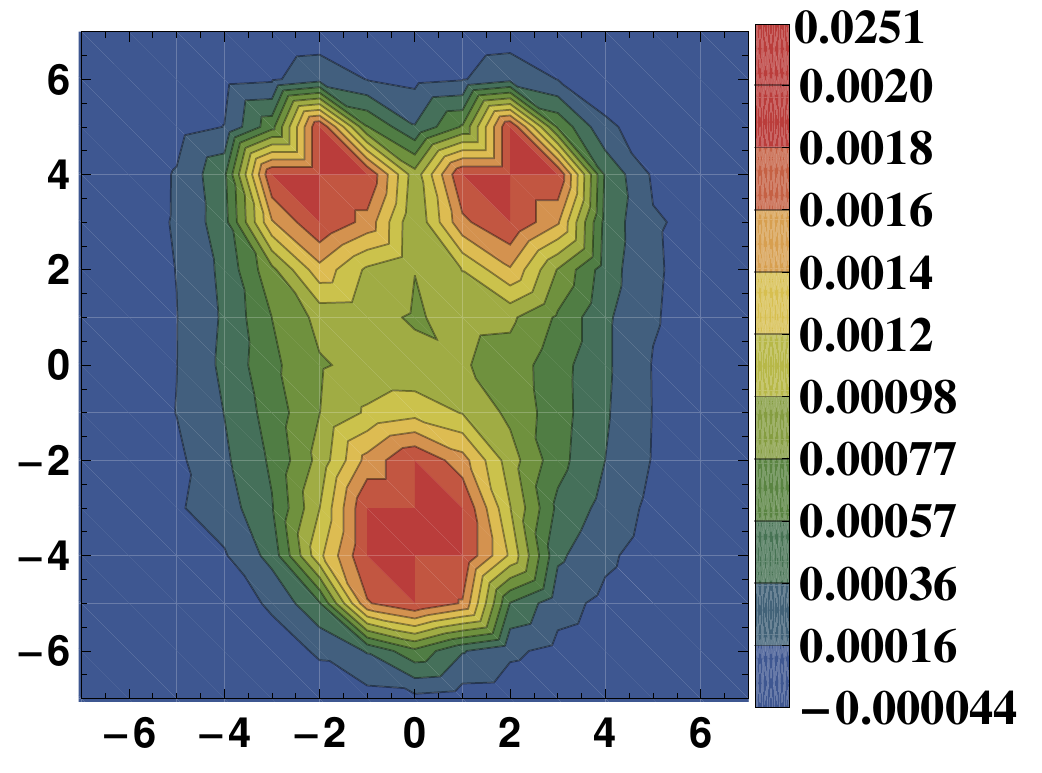}
\par\end{centering}}
\par\end{centering}
    \caption{Results for the U geometry with $d=4$ and $l=8$}
    \label{cfield_U_d_4_l_8}
\end{figure}

\begin{figure}[H]
\begin{centering}
    \subfloat[Chromoelectric Field\label{fig:cfield_U_d_6_l_8_E}]{
\begin{centering}
    \includegraphics[width=3.5cm]{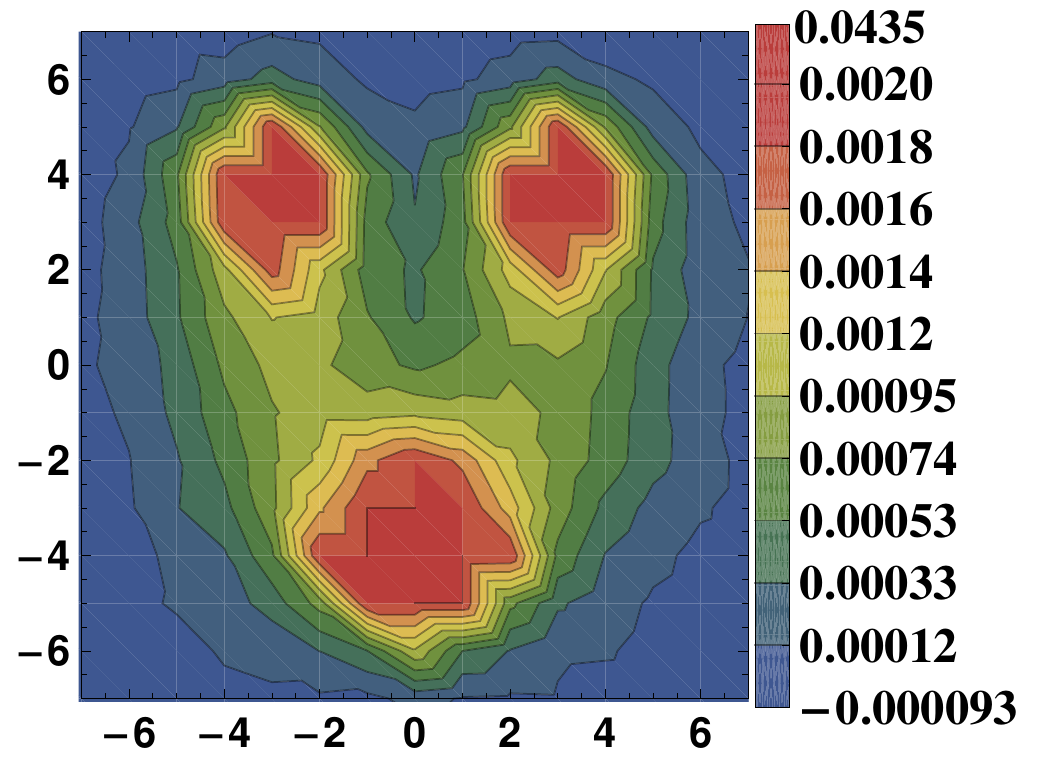}
\par\end{centering}}
    \subfloat[Chromomagnetic Field\label{fig:cfield_U_d_6_l_8_B}]{
\begin{centering}
    \includegraphics[width=3.5cm]{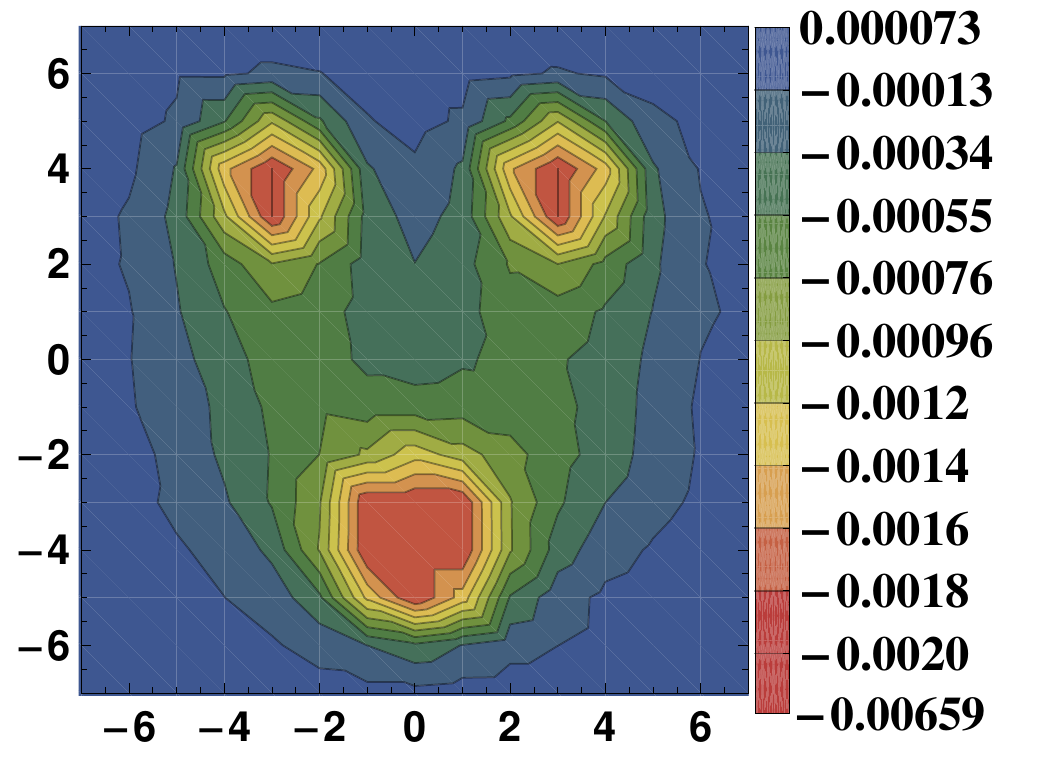}
\par\end{centering}}
    \subfloat[Energy Density\label{fig:cfield_U_d_6_l_8_Energ}]{
\begin{centering}
    \includegraphics[width=3.5cm]{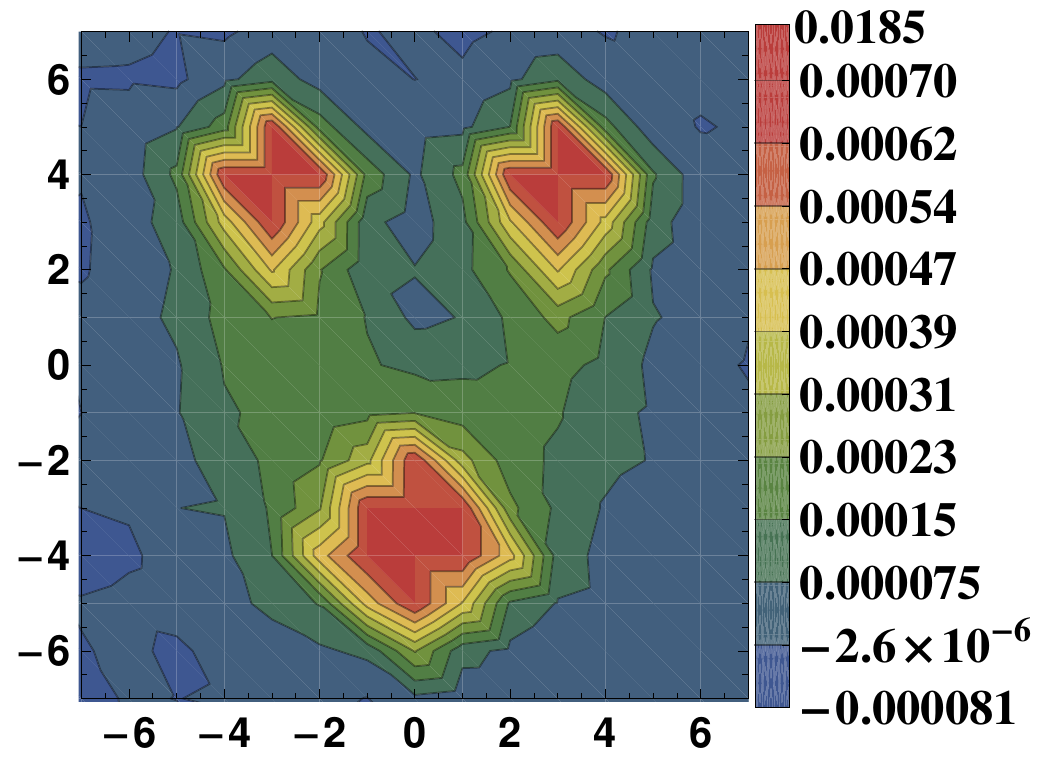}
\par\end{centering}}
    \subfloat[Action Density\label{fig:cfield_U_d_6_l_8_Act}]{
\begin{centering}
    \includegraphics[width=3.5cm]{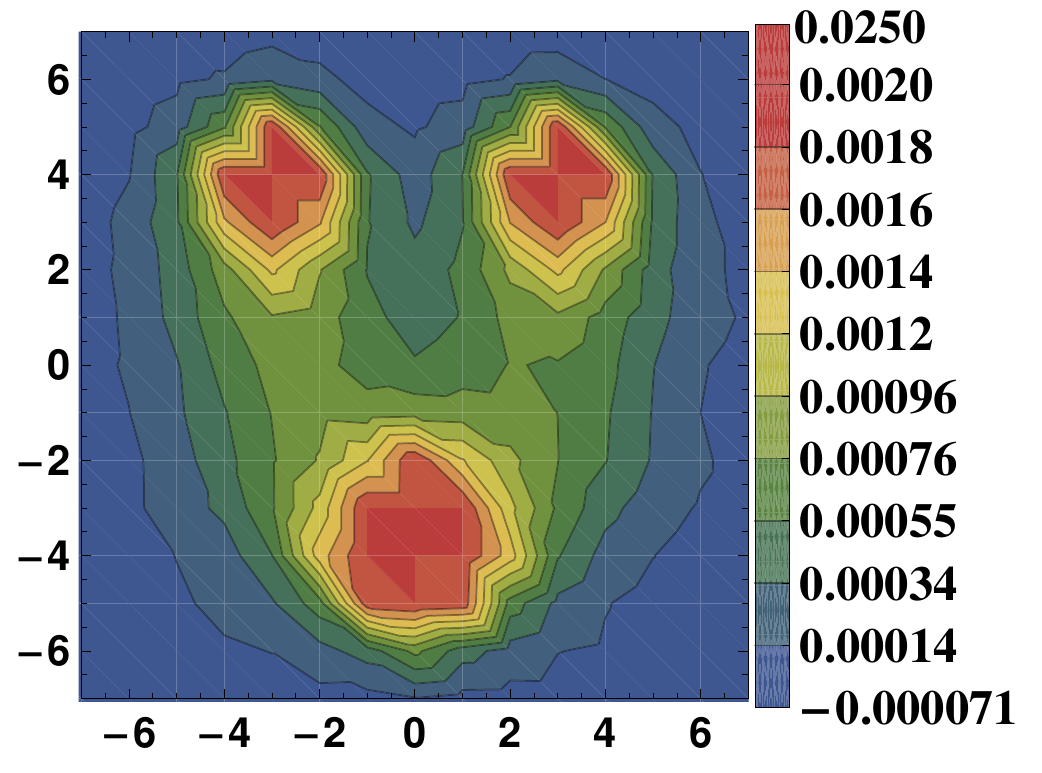}
\par\end{centering}}
\par\end{centering}
    \caption{Results for the U geometry with $d=6$ and $l=8$}
    \label{cfield_U_d_6_l_8}
\end{figure}

\begin{figure}[H]
\begin{centering}
    \subfloat[Chromoelectric Field\label{fig:cfield_qqg_L_r1_0_r2_8_E}]{
\begin{centering}
    \includegraphics[width=3.5cm]{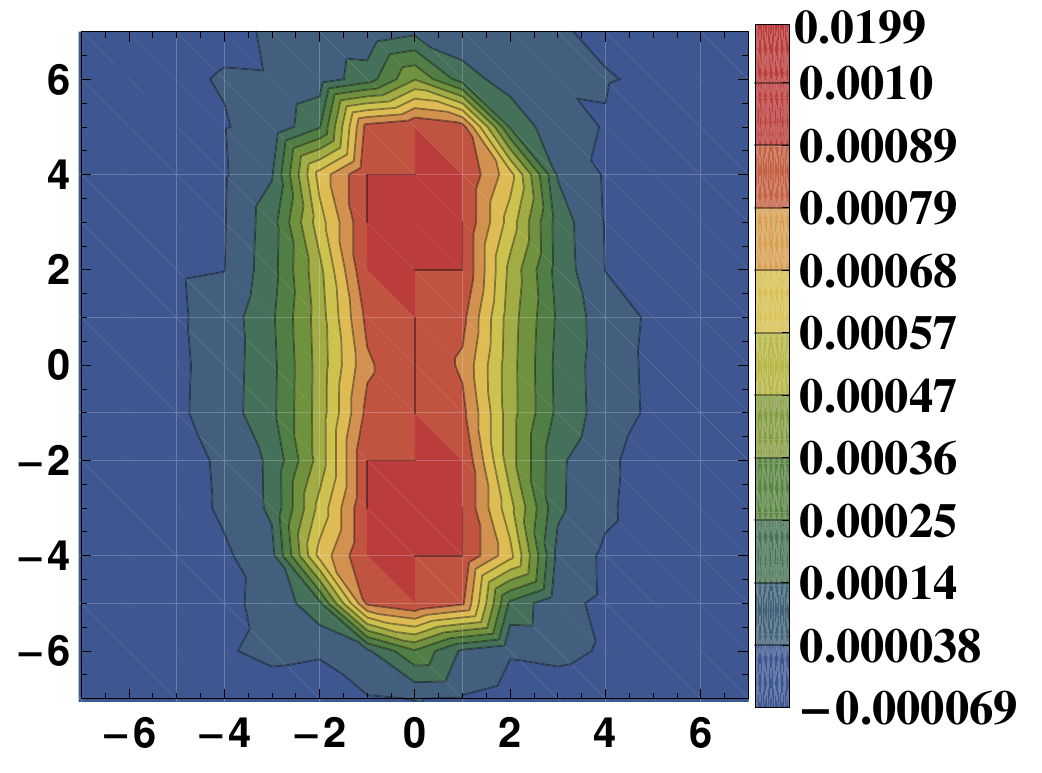}
\par\end{centering}}
    \subfloat[Chromomagnetic Field\label{fig:cfield_qqg_L_r1_0_r2_8_B}]{
\begin{centering}
    \includegraphics[width=3.5cm]{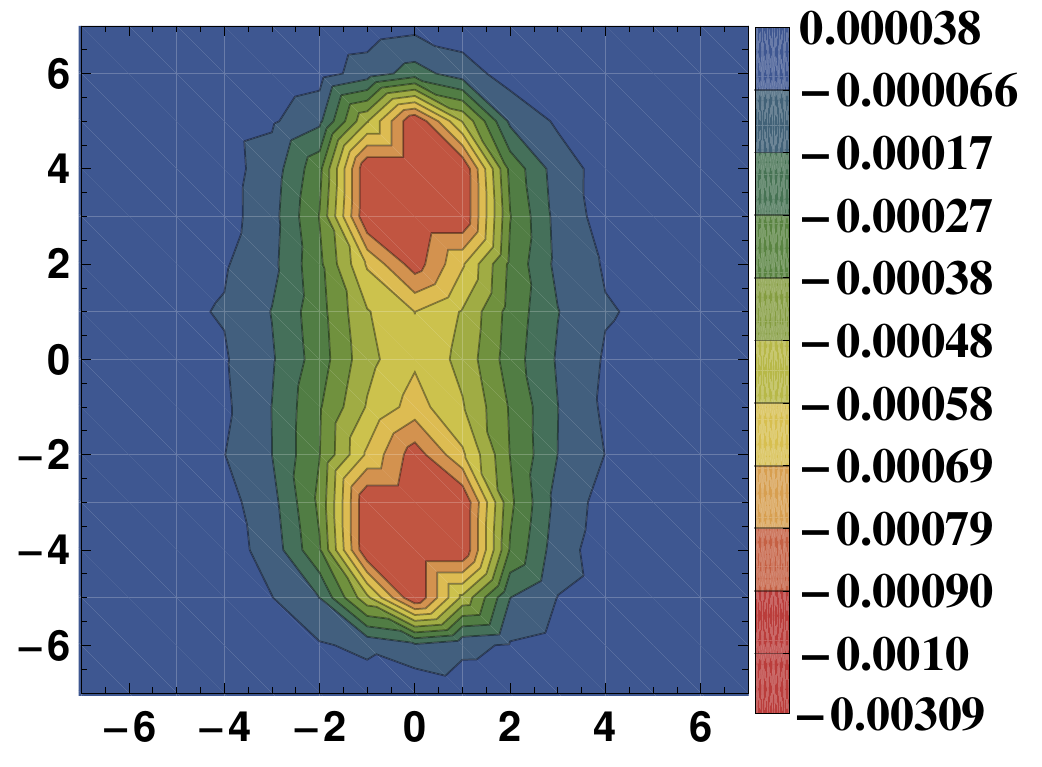}
\par\end{centering}}
    \subfloat[Energy Density\label{fig:cfield_qqg_L_r1_0_r2_8_Energ}]{
\begin{centering}
    \includegraphics[width=3.5cm]{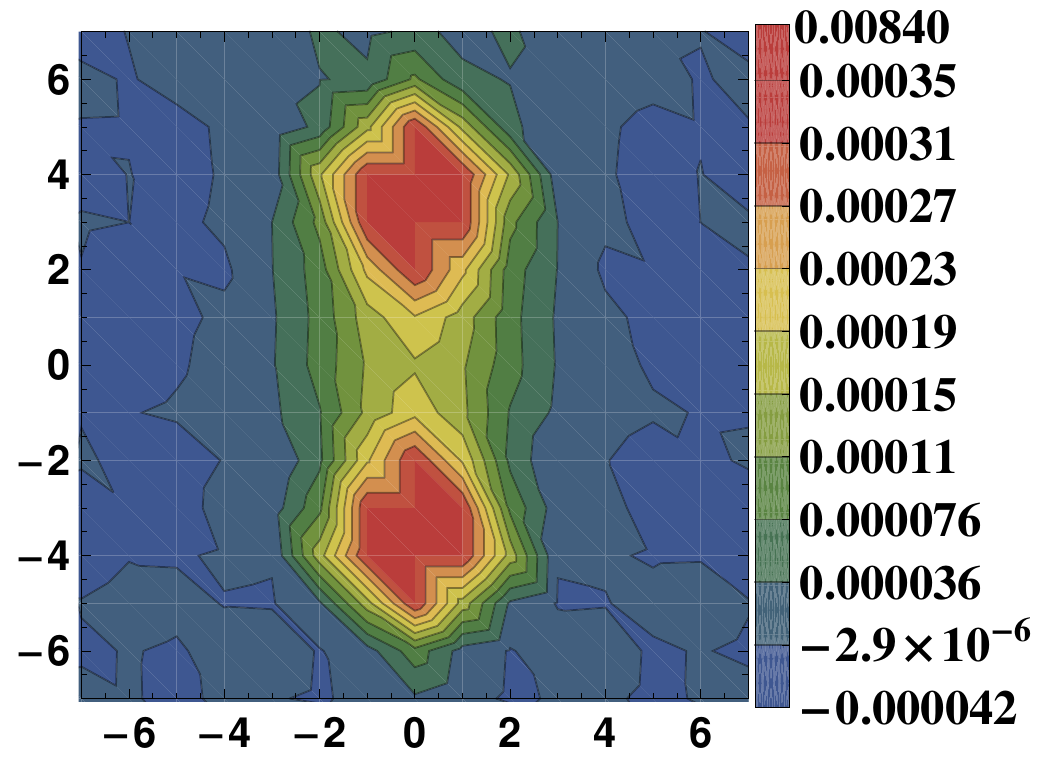}
\par\end{centering}}
    \subfloat[Action Density\label{fig:cfield_qqg_L_r1_0_r2_8_Act}]{
\begin{centering}
    \includegraphics[width=3.5cm]{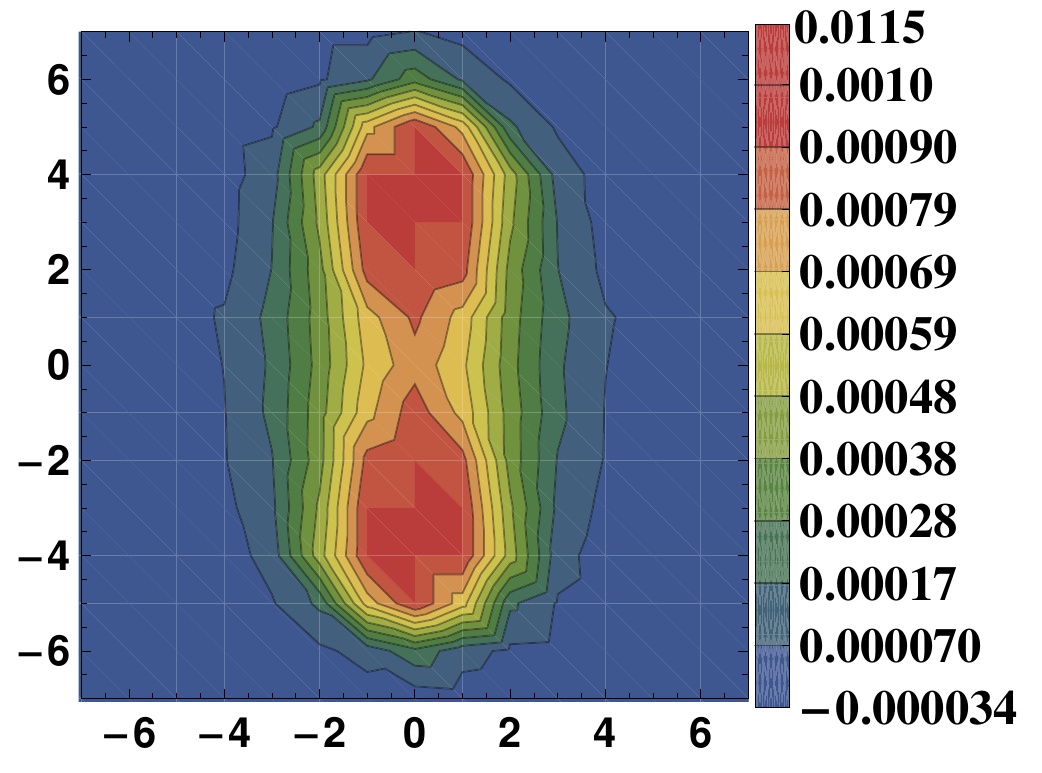}
\par\end{centering}}
\par\end{centering}
    \caption{Results for the L geometry with $r_1=0$ and $r_2=8$. The gluon and the antiquark
are superposed and the results are consistent with the static quark-antiquark case.}
    \label{cfield_qqg_L_r1_0_r2_8}
\end{figure}

\begin{figure}[H]
\begin{centering}
    \subfloat[Chromoelectric Field\label{fig:cfield_qqg_L_r1_2_r2_8_E}]{
\begin{centering}
    \includegraphics[width=3.5cm]{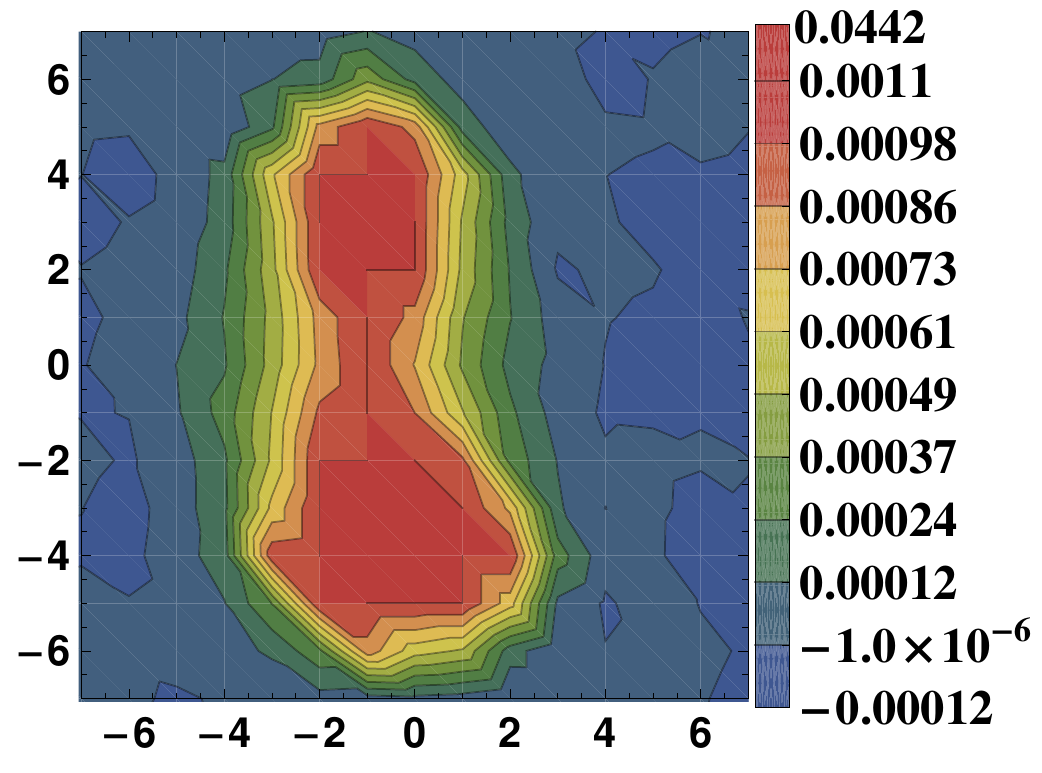}
\par\end{centering}}
    \subfloat[Chromomagnetic Field\label{fig:cfield_qqg_L_r1_2_r2_8_B}]{
\begin{centering}
    \includegraphics[width=3.5cm]{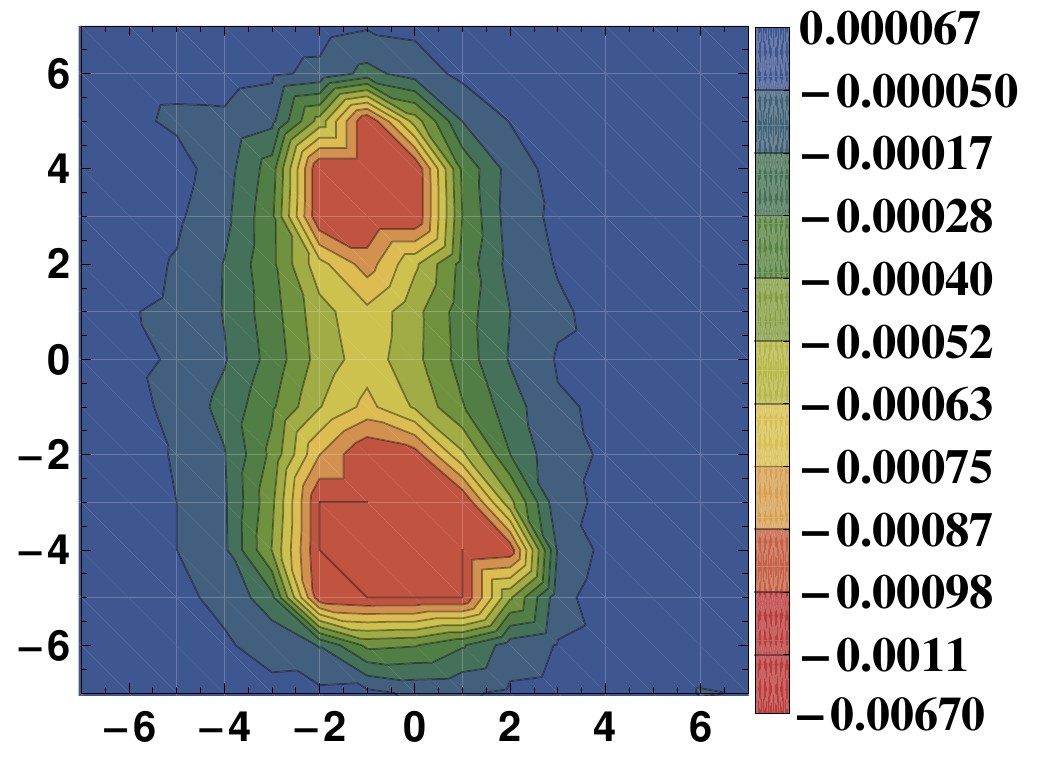}
\par\end{centering}}
    \subfloat[Energy Density\label{fig:cfield_qqg_L_r1_2_r2_8_Energ}]{
\begin{centering}
    \includegraphics[width=3.5cm]{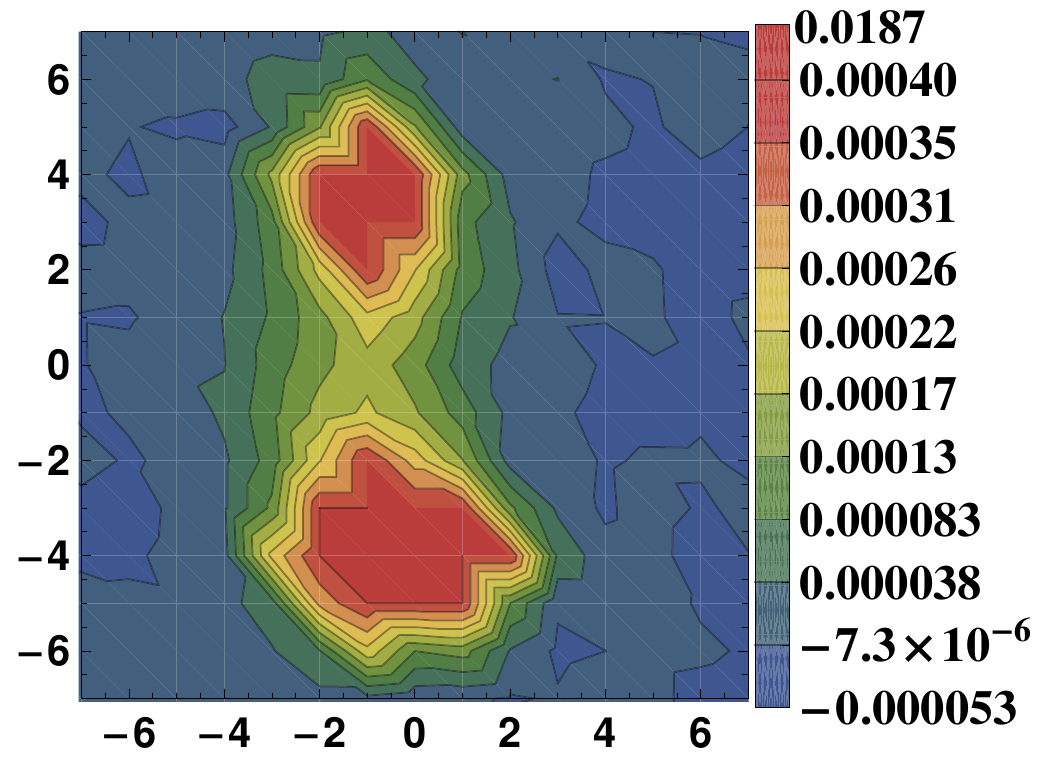}
\par\end{centering}}
    \subfloat[Action Density\label{fig:cfield_qqg_L_r1_2_r2_8_Act}]{
\begin{centering}
    \includegraphics[width=3.5cm]{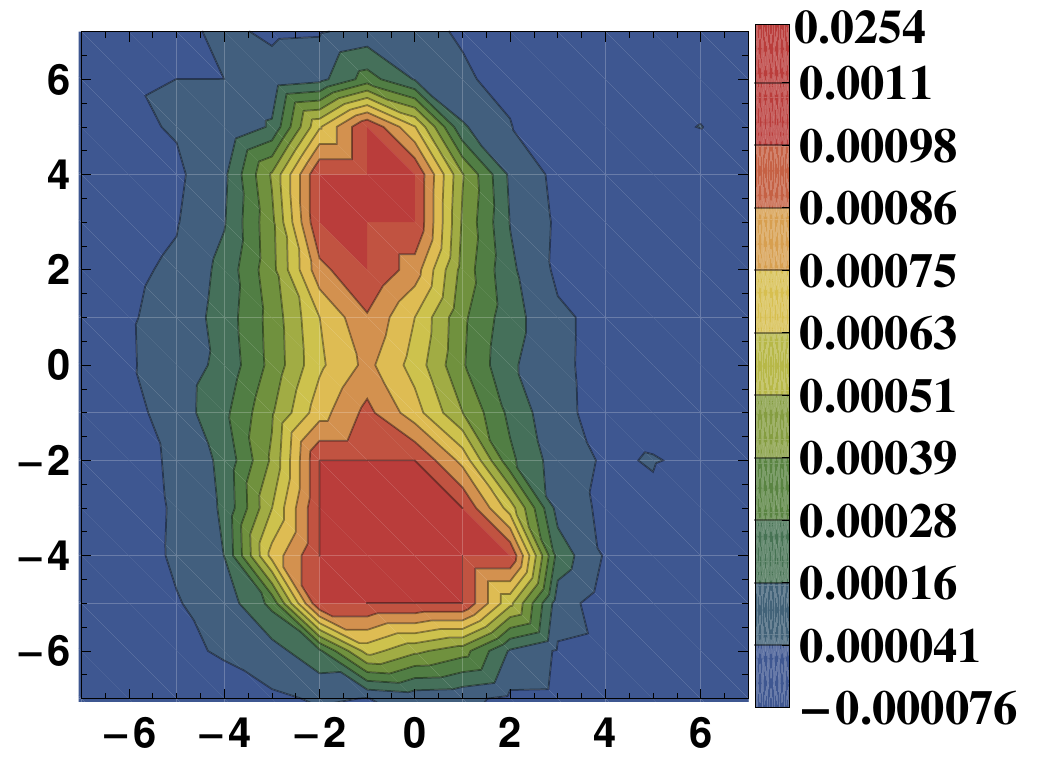}
\par\end{centering}}
\par\end{centering}
    \caption{Results for the L geometry with $r_1=2$ and $r_2=8$}
    \label{cfield_qqg_L_r1_2_r2_8}
\end{figure}

\begin{figure}[H]
\begin{centering}
    \subfloat[Chromoelectric Field\label{fig:cfield_qqg_L_r1_8_r2_8_E}]{
\begin{centering}
    \includegraphics[width=3.5cm]{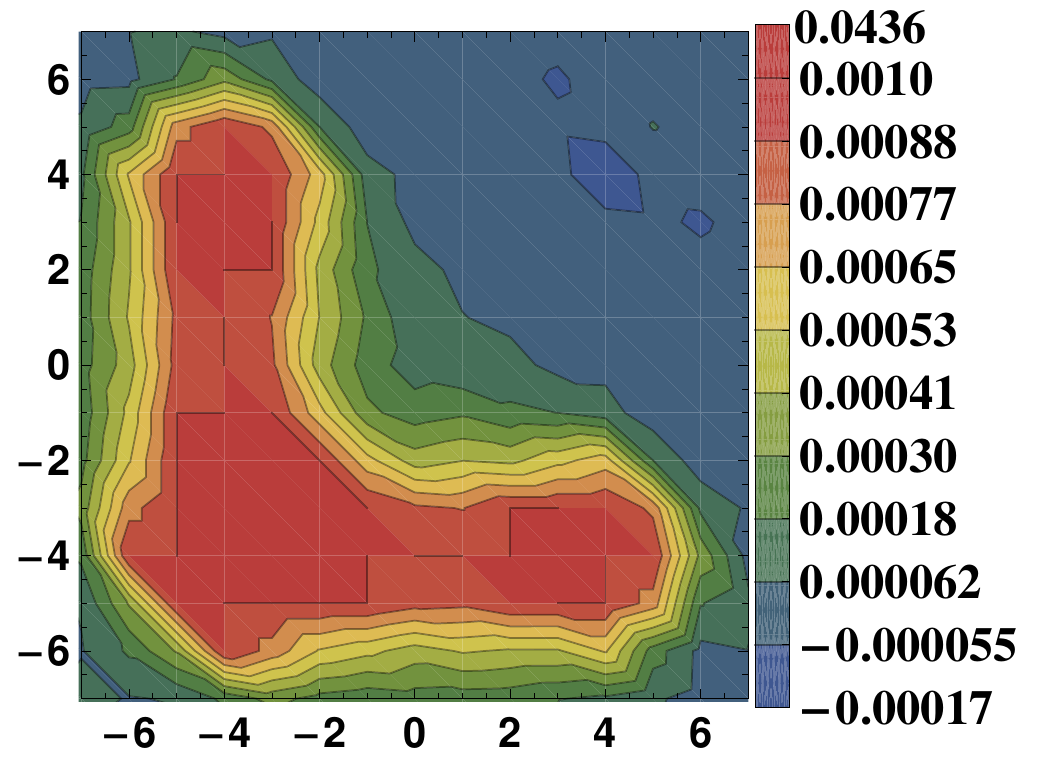}
\par\end{centering}}
    \subfloat[Chromomagnetic Field\label{fig:cfield_qqg_L_r1_8_r2_8_B}]{
\begin{centering}
    \includegraphics[width=3.5cm]{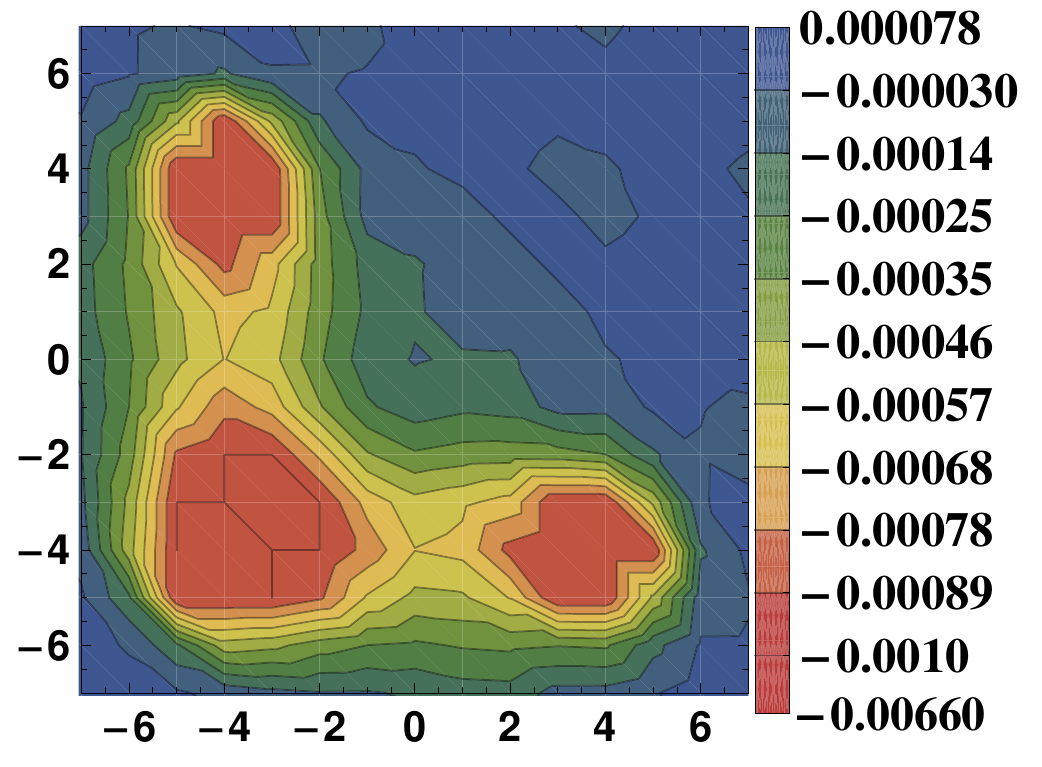}
\par\end{centering}}
    \subfloat[Energy Density\label{fig:cfield_qqg_L_r1_8_r2_8_Energ}]{
\begin{centering}
    \includegraphics[width=3.5cm]{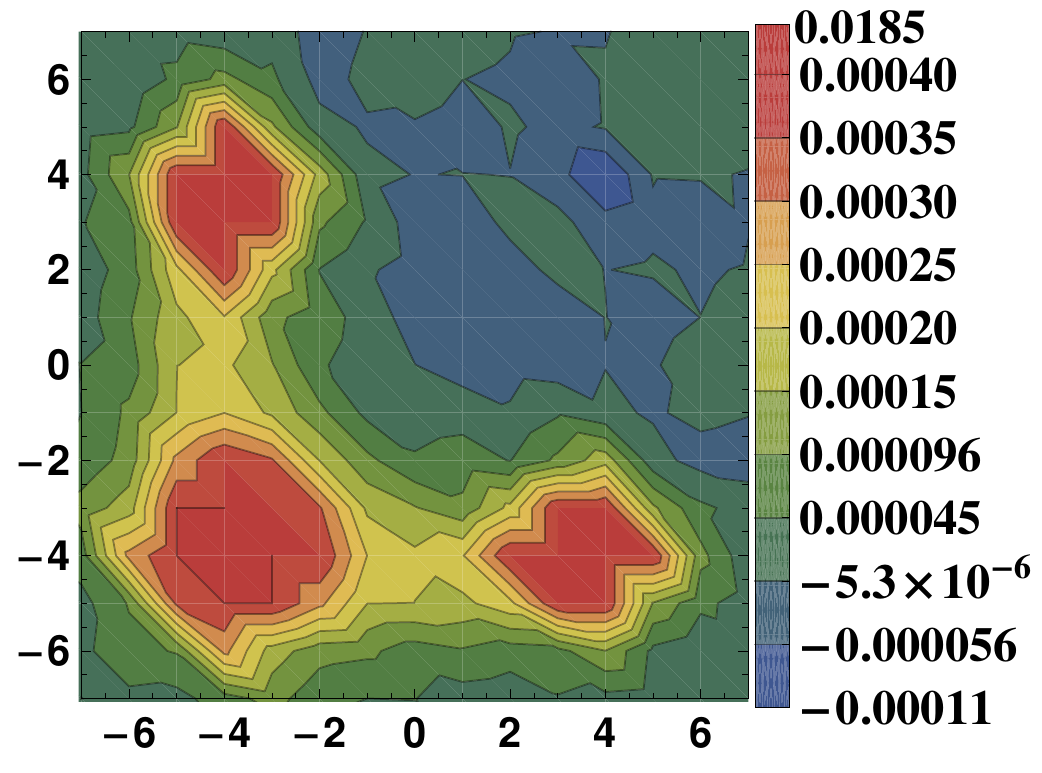}
\par\end{centering}}
    \subfloat[Action Density\label{fig:cfield_qqg_L_r1_8_r2_8_Act}]{
\begin{centering}
    \includegraphics[width=3.5cm]{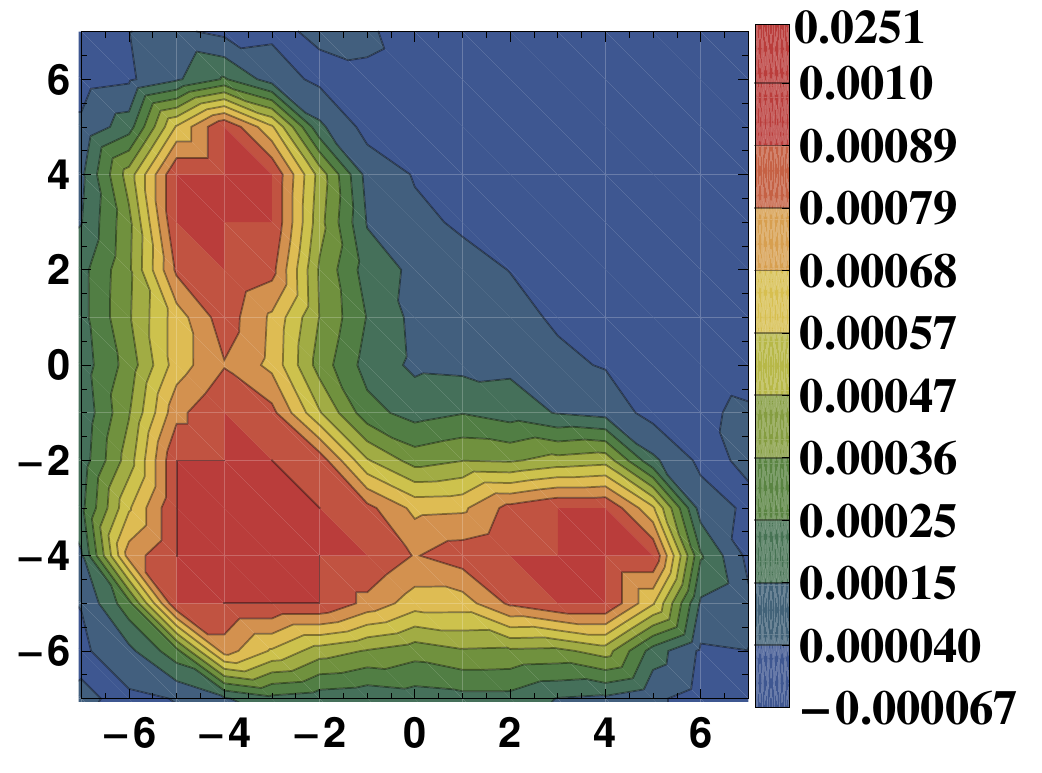}
\par\end{centering}}
\par\end{centering}
    \caption{Results for the L geometry with $r_1=8$ and $r_2=8$}
    \label{cfield_qqg_L_r1_8_r2_8}
\end{figure}

\begin{figure}[H]
\begin{centering}
    \subfloat[Chromoelectric Field\label{fig:cfield_qqg_L_r1_6_r2_6_E}]{
\begin{centering}
    \includegraphics[width=3.5cm]{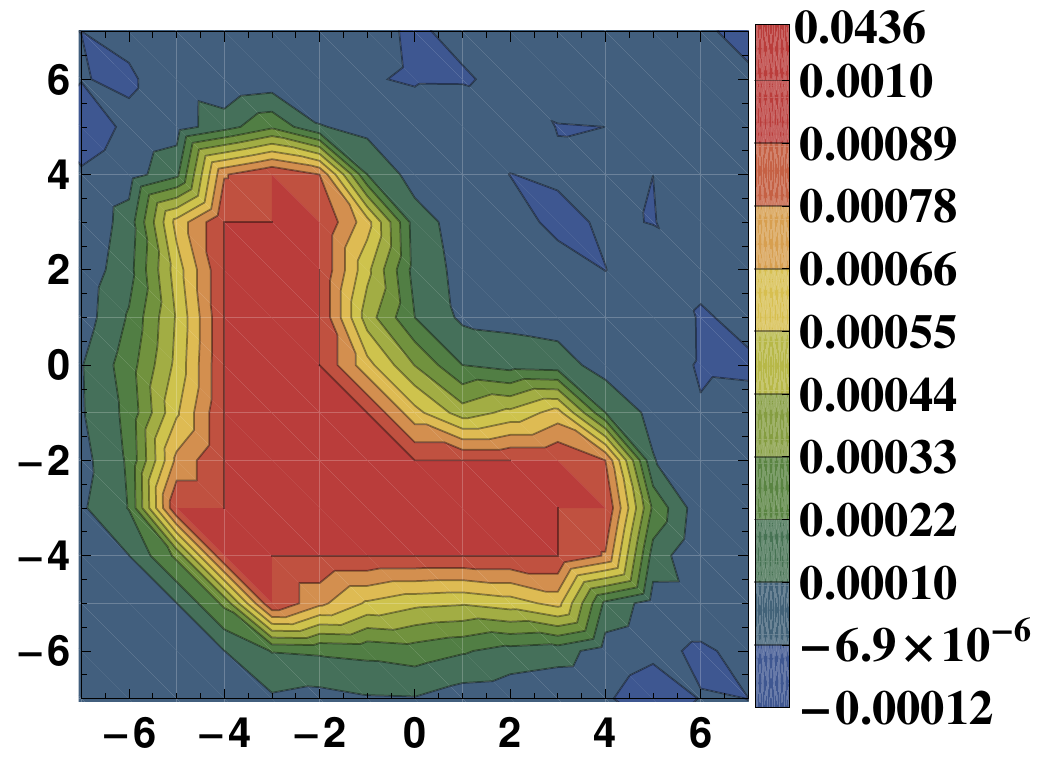}
\par\end{centering}}
    \subfloat[Chromomagnetic Field\label{fig:cfield_qqg_L_r1_6_r2_6_B}]{
\begin{centering}
    \includegraphics[width=3.5cm]{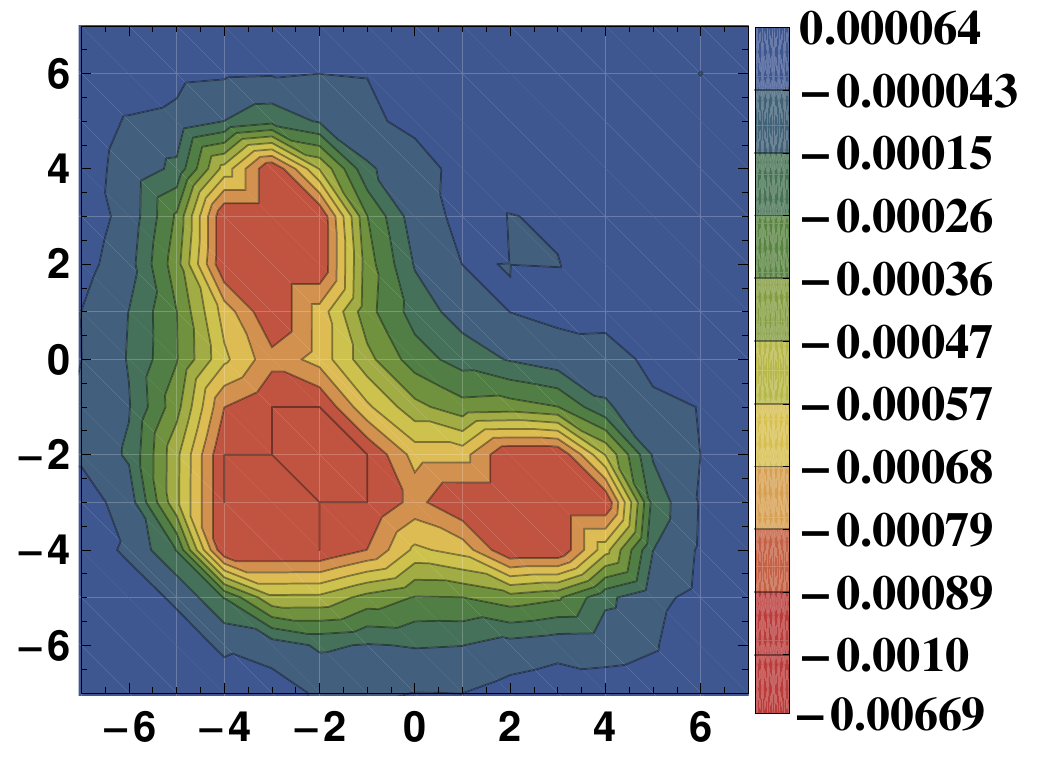}
\par\end{centering}}
    \subfloat[Energy Density\label{fig:cfield_qqg_L_r1_6_r2_6_Energ}]{
\begin{centering}
    \includegraphics[width=3.5cm]{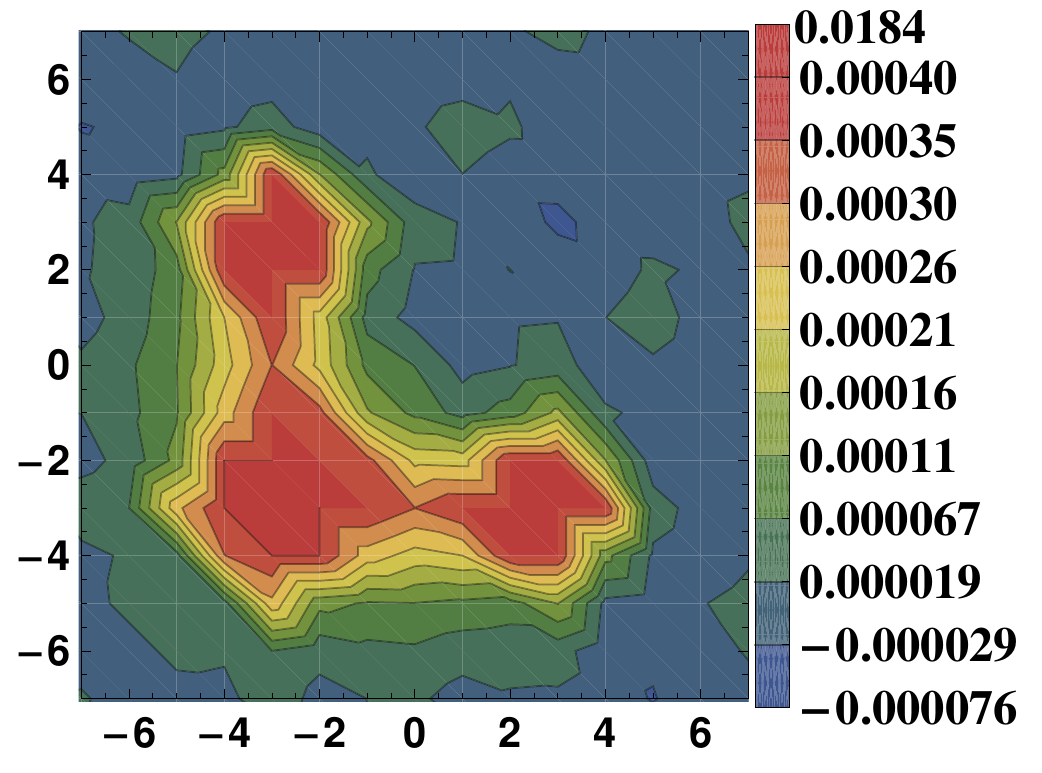}
\par\end{centering}}
    \subfloat[Action Density\label{fig:cfield_qqg_L_r1_6_r2_6_Act}]{
\begin{centering}
    \includegraphics[width=3.5cm]{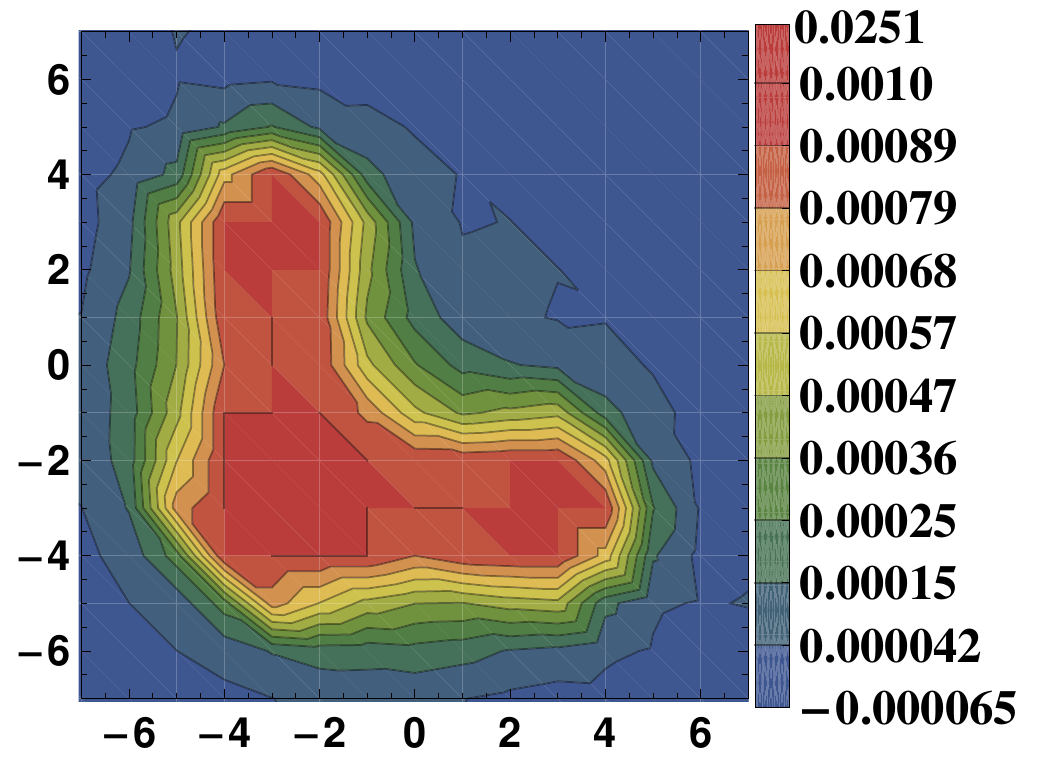}
\par\end{centering}}
\par\end{centering}
    \caption{Results for the L geometry with $r_1=6$ and $r_2=6$}
    \label{cfield_qqg_L_r1_6_r2_6}
\end{figure}

In the following pictures we present the results for the chromolectric and chromomagnetic fields and for the energy and action density. The figures \ref{cfield_U_d_0_l_8} to \ref{cfield_U_d_6_l_8} show the results for the U shape geometry and figures \ref{cfield_qqg_L_r1_0_r2_8} to \ref{cfield_qqg_L_r1_6_r2_6} the results for the L shape geometry.

\begin{figure}[H]
\begin{centering}

    \subfloat[\label{fig:cfield_U_d_0_l_8_All_2D}$d=0$ and $l=8$ for $x=0$]{
\begin{centering}
    \includegraphics[height=2.2cm]{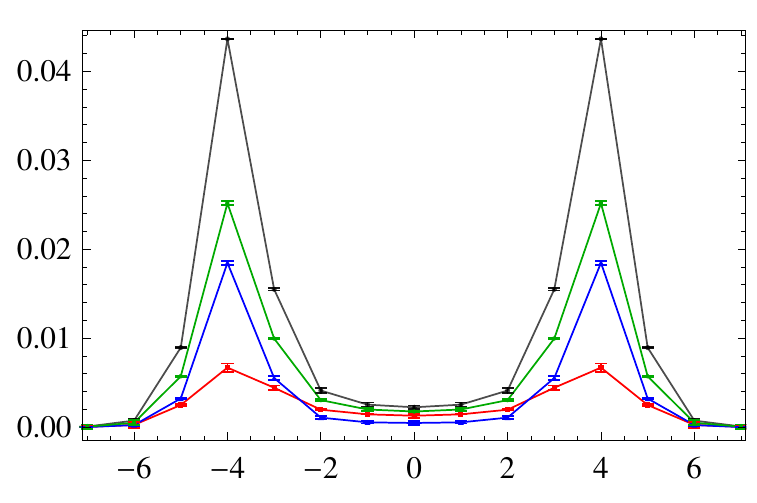}
\par\end{centering}}
    \subfloat[\label{fig:cfield_U_d_0_l_8_All0_2D}$d=0$ and $l=8$ for $y=4$]{
\begin{centering}
    \includegraphics[height=2.2cm]{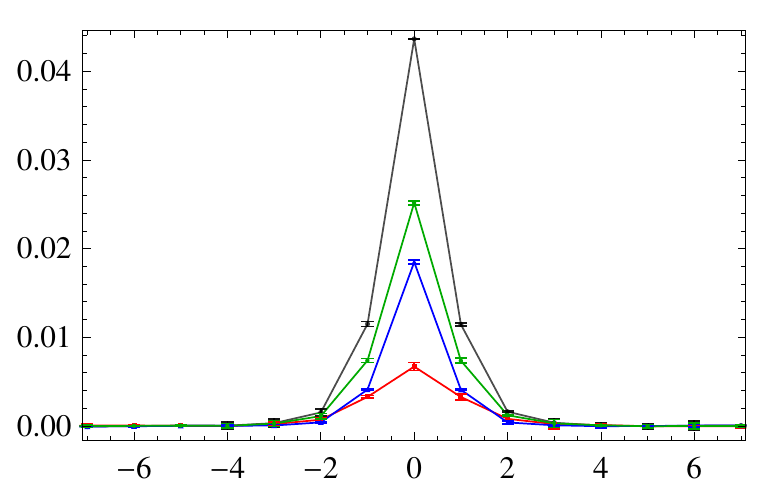}
\par\end{centering}}
    \subfloat[\label{fig:cfield_U_d_2_l_8_All_2D}$d=2$ and $l=8$ for $x=0$]{
\begin{centering}
    \includegraphics[height=2.2cm]{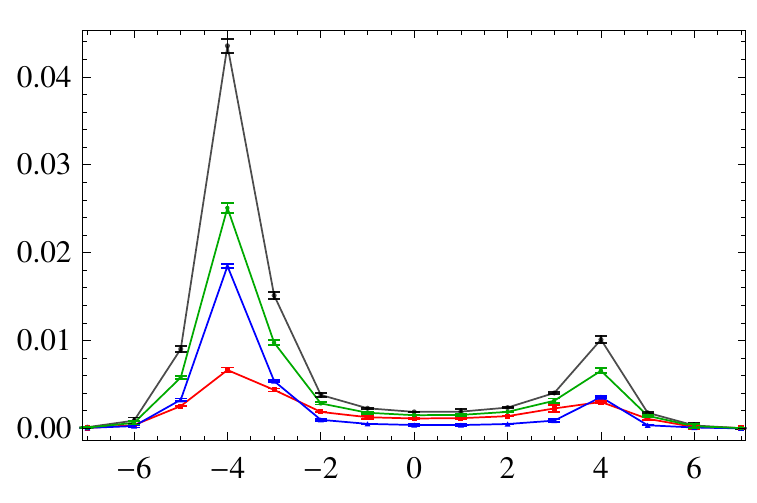}
\par\end{centering}}
    \subfloat[\label{fig:cfield_U_d_2_l_8_All0_2D}$d=2$ and $l=8$ for $y=4$]{
\begin{centering}
    \includegraphics[height=2.2cm]{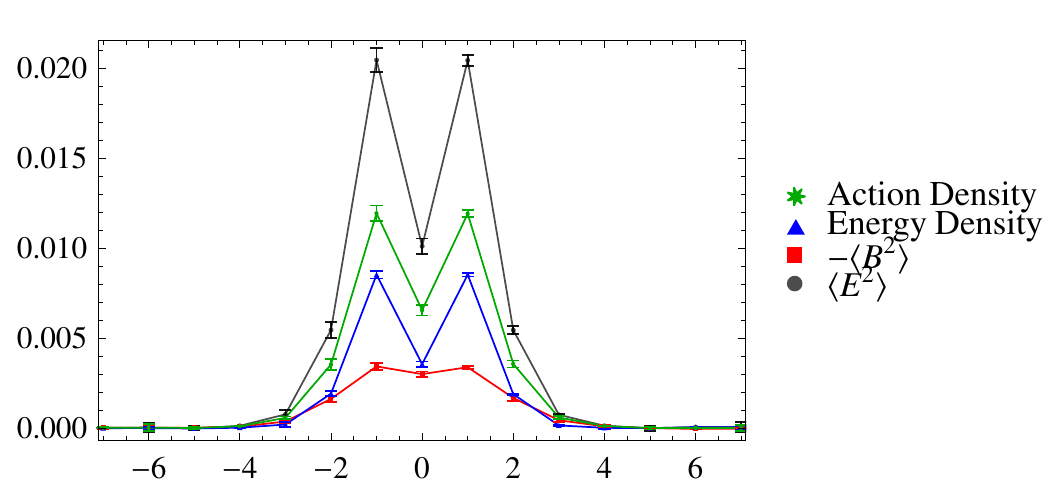}
\par\end{centering}}

    \subfloat[\label{fig:cfield_U_d_4_l_8_All_2D}$d=4$ and $l=8$ for $x=0$]{
\begin{centering}
    \includegraphics[height=2.2cm]{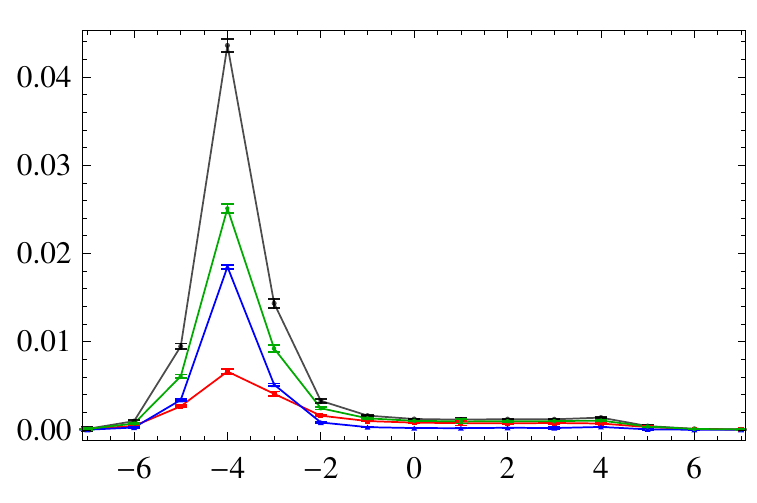}
\par\end{centering}}
    \subfloat[\label{fig:cfield_U_d_4_l_8_All0_2D}$d=4$ and $l=8$ for $y=4$]{
\begin{centering}
    \includegraphics[height=2.2cm]{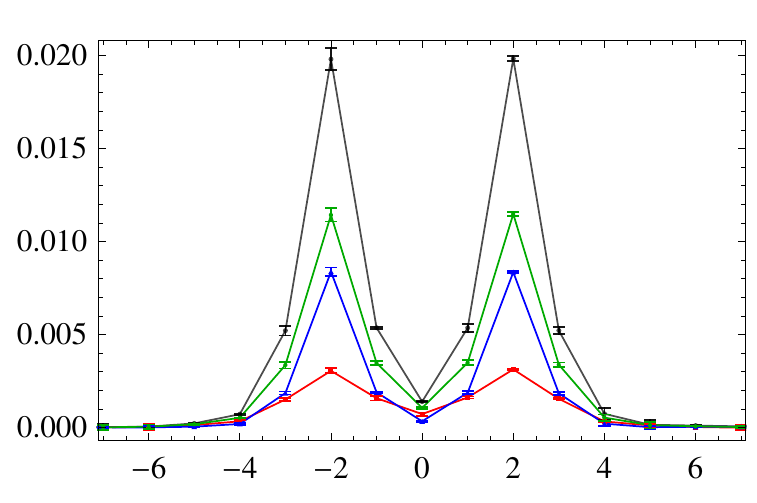}
\par\end{centering}}
    \subfloat[\label{fig:cfield_U_d_6_l_8_All_2D}$d=6$ and $l=8$ for $x=0$]{
\begin{centering}
    \includegraphics[height=2.2cm]{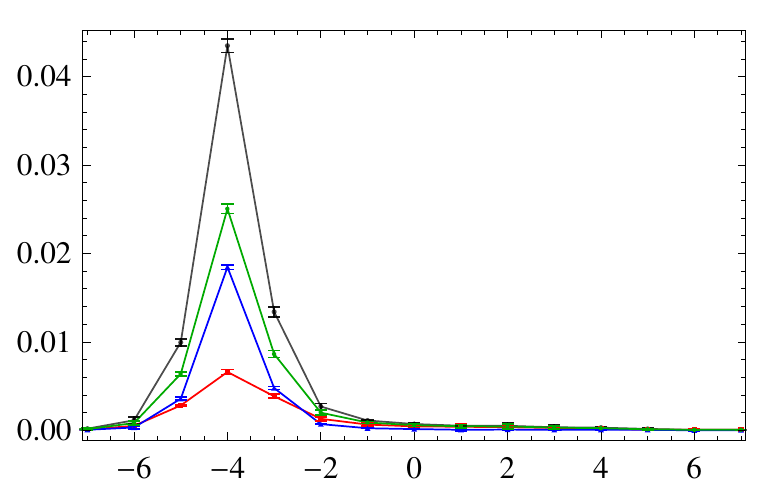}
\par\end{centering}}
    \subfloat[\label{fig:cfield_U_d_6_l_8_All0_2D}$d=6$ and $l=8$ for $y=4$]{
\begin{centering}
    \includegraphics[height=2.2cm]{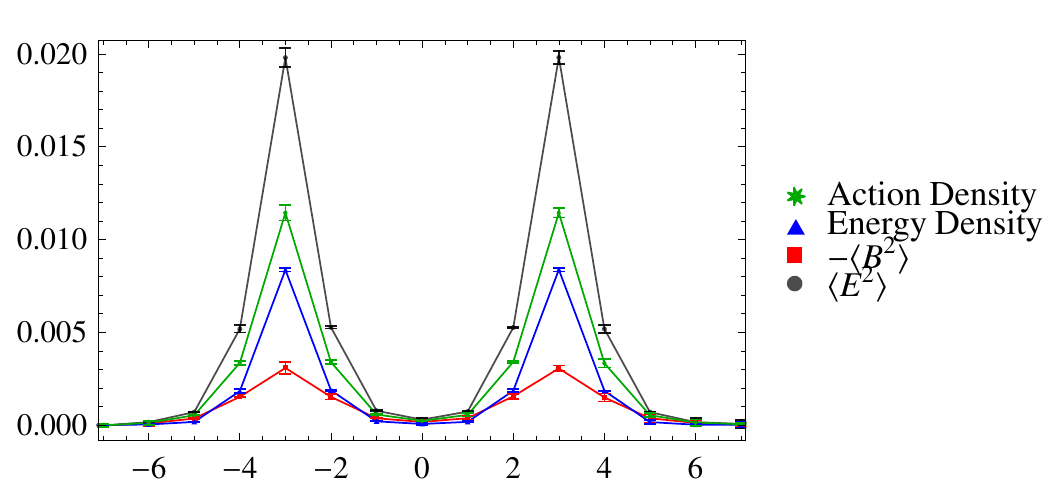}
\par\end{centering}}

\par\end{centering}

    \caption{Results for the U geometry.}
    \label{cfield_U_All}
\end{figure}

\begin{figure}[H]
\begin{centering}

    \subfloat[\label{fig:cfield_qqg_L_d_0_l_8_Allx_2D}$r_1=0$ and $r_2=8$, along segment gluon-antiquark]{
\begin{centering}
    \includegraphics[height=2.2cm]{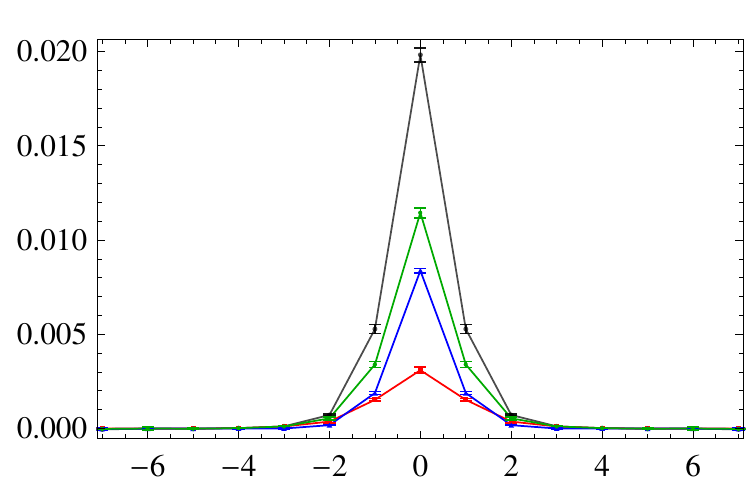}
\par\end{centering}}
    \subfloat[\label{fig:cfield_qqg_L_r1_0_r2_8_Ally_2D}$r_1=0$ and $r_2=8$, along segment gluon-quark]{
\begin{centering}
    \includegraphics[height=2.2cm]{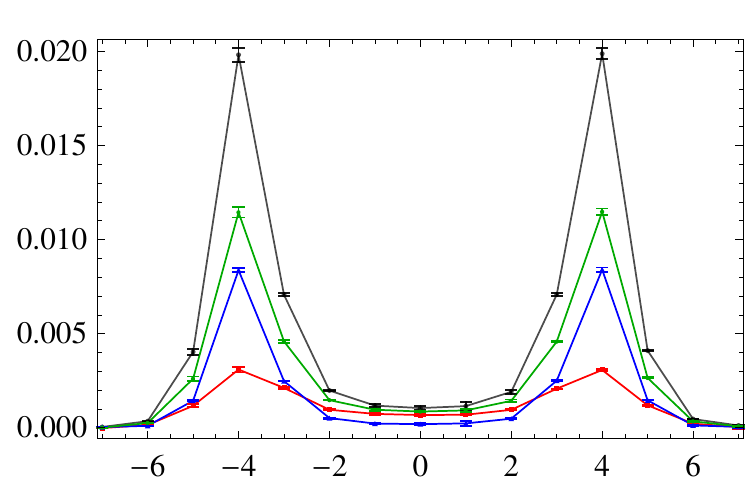}
\par\end{centering}}
    \subfloat[\label{fig:cfield_qqg_L_r1_2_r2_8_Allx_2D}$r_1=2$ and $r_2=8$, along segment gluon-antiquark]{
\begin{centering}
    \includegraphics[height=2.2cm]{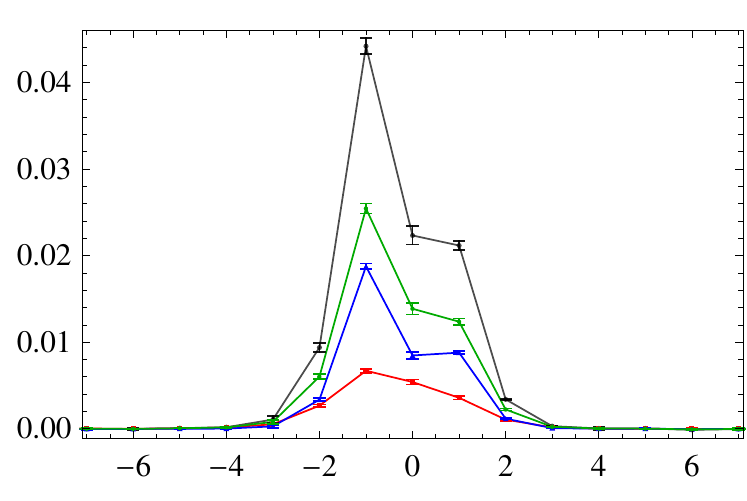}
\par\end{centering}}
    \subfloat[\label{fig:cfield_qqg_L_r1_2_r2_8_Ally_2D}$r_1=2$ and $r_2=8$, along segment gluon-quark]{
\begin{centering}
    \includegraphics[height=2.2cm]{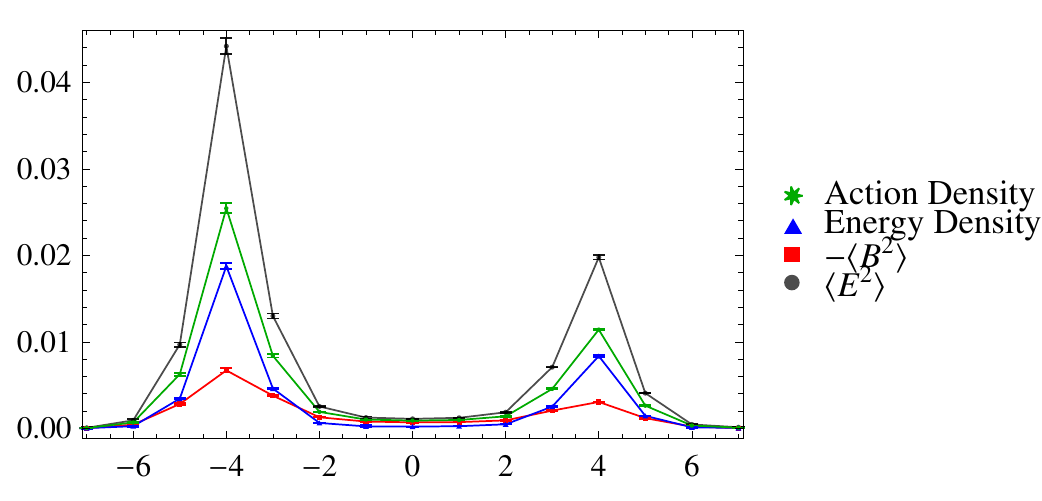}
\par\end{centering}}

    \subfloat[\label{fig:cfield_qqg_L_r1_8_r2_8_Allx_2D}$r_1=8$ and $r_2=8$, along segment gluon-antiquark]{
\begin{centering}
    \includegraphics[height=2.2cm]{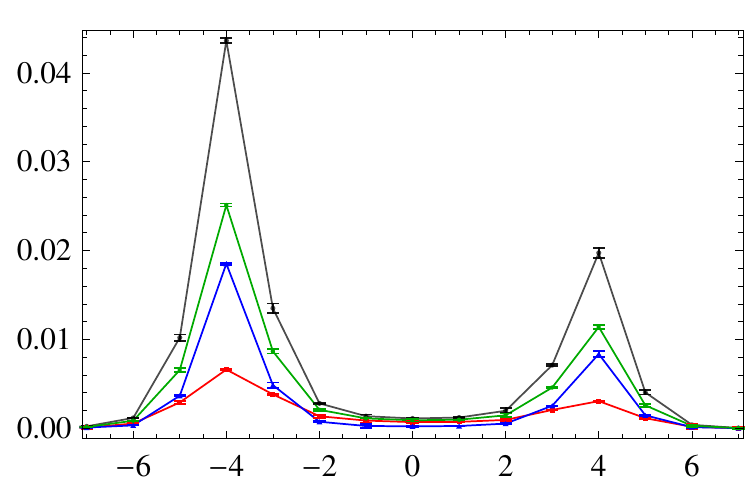}
\par\end{centering}}
    \subfloat[\label{fig:cfield_qqg_L_d_4_l_8_Ally_2D}$r_1=8$ and $r_2=8$, along segment gluon-quark]{
\begin{centering}
    \includegraphics[height=2.2cm]{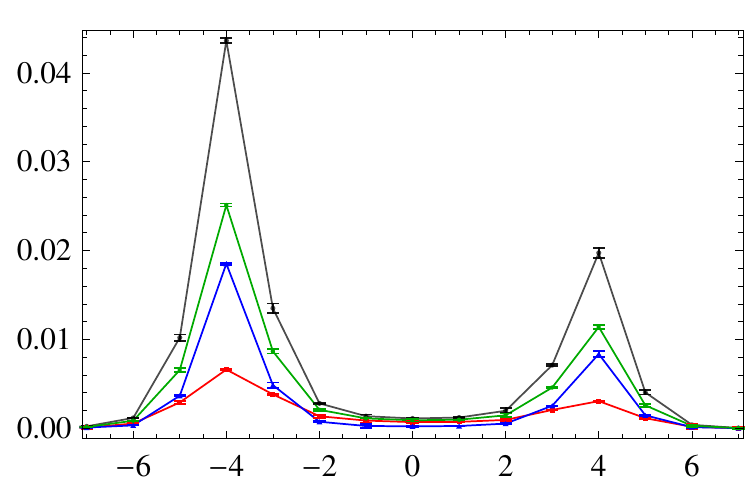}
\par\end{centering}}
    \subfloat[\label{fig:cfield_qqg_L_r1_6_r2_6_Allx_2D}$r_1=6$ and $r_2=6$, along segment gluon-antiquark]{
\begin{centering}
    \includegraphics[height=2.2cm]{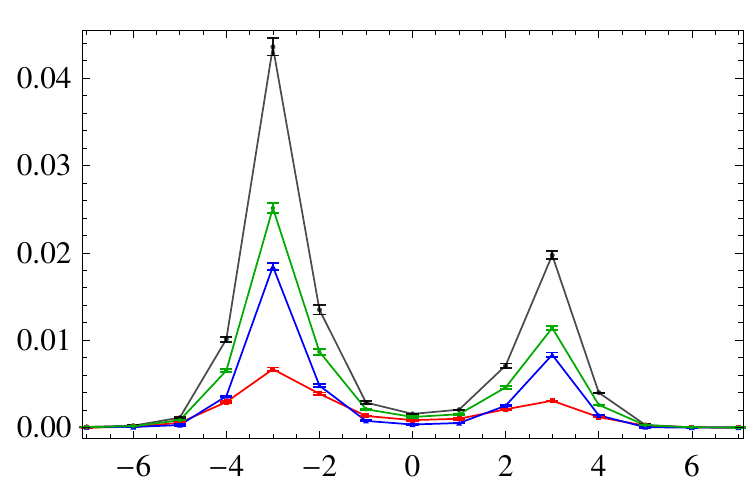}
\par\end{centering}}
    \subfloat[\label{fig:cfield_qqg_L_r1_6_r2_6_Ally_2D}$r_1=6$ and $r_2=6$, along segment gluon-quark]{
\begin{centering}
    \includegraphics[height=2.2cm]{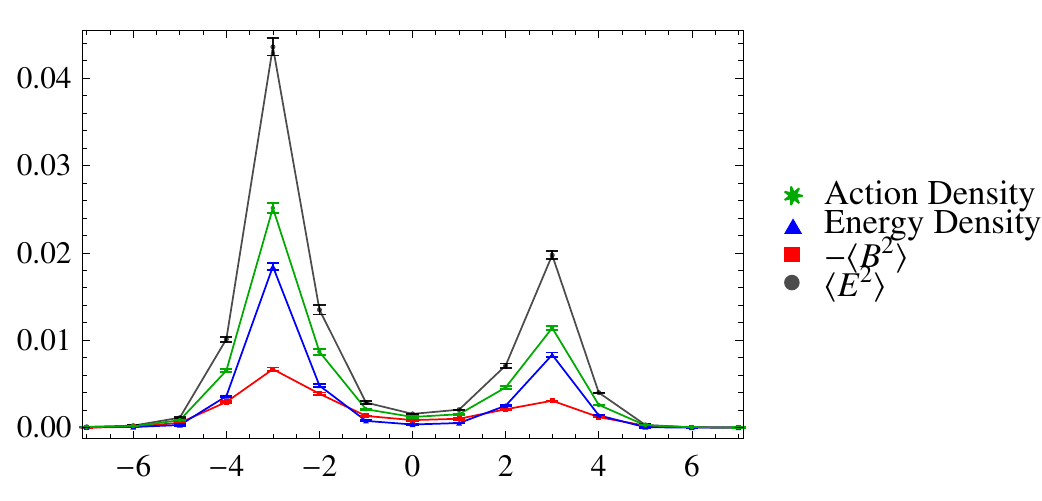}
\par\end{centering}}

\par\end{centering}

    \caption{Results for the L geometry.}
    \label{cfield_qqg_L_All}
\end{figure}

\begin{figure}[H]
\begin{center}
    \includegraphics[width=12cm]{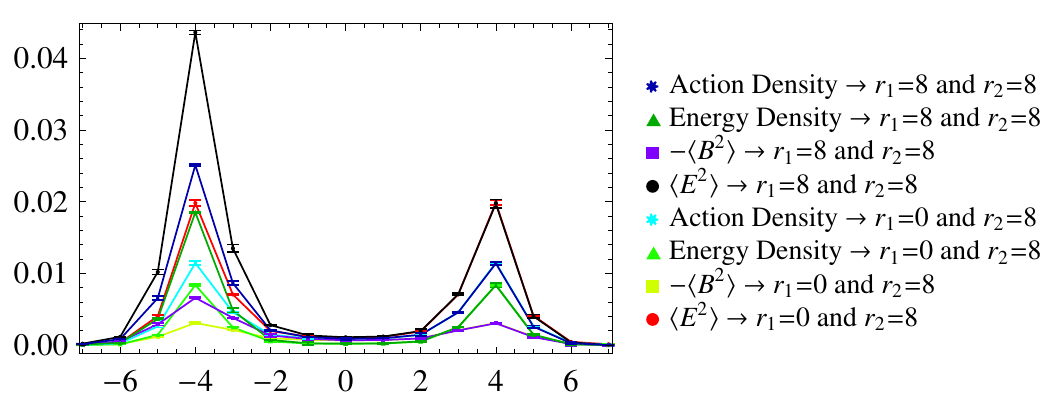}
    \caption{Comparison between the results for ($r_1 = 0; r_2 = 8$) and ($r_1 = 8; r_2 = 8$) along segment gluon-antiquark in the L geometry.}
    \label{qq_qqg}
\end{center}
\end{figure}

In figure \ref{cfield_U_All} we present the results for the U geometry along $x=0$, this corresponds to the axis along gluon and the middle distance between quark-antiquark, and $y=4$, this corresponds to the axis along quark-antiquark. This results show that when the quark and the antiquark are superposed the results are consistent with the degenerate case of the two gluon glueball.
In figure \ref{cfield_qqg_L_All} we present the results for the L geometry along the segment gluon-antiquark and along segment gluon-quark. And finally, in figure \ref{qq_qqg} we show a comparison between the results for $(r_1=0,r_2=8)$ and $(r_1=8,r_2=8)$ along segment gluon-antiquark in the L geometry, when the gluon and the antiquark are superposed this results are consistent with the static quark-antiquark case.

\begin{figure}[h]
\begin{center}
    \includegraphics[width=7cm]{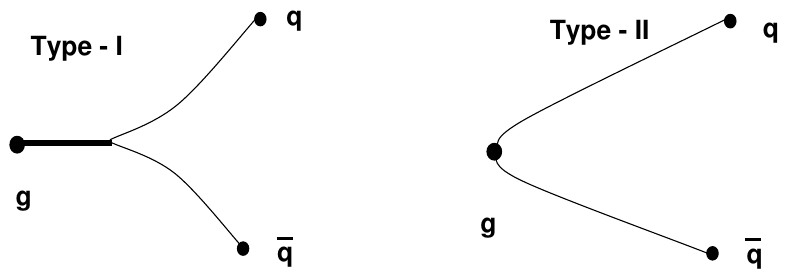}
    \caption{String attraction and fusion, and string repulsion, respectively in type I and II superconductors.}
    \label{superconductor}
\end{center}
\end{figure}

\section{Conclusions}
In Bicudo et al. \cite{Bicudo2008a} and Cardoso et al. \cite{Cardoso2007} it was shown that
for the static gluon-quark-antiquark potential $V$ when the segments between gluon-quark and gluon-antiquark are perpendicular, $V$ is compatible with confinement realized with a pair of fundamental strings, one linking the gluon to the quark and the other linking the same gluon to the antiquark.
However, if the segments are parallel and superposed, the total string tension becomes larger and agrees with Casimir Scaling measured by Bali \cite{Bali2000}. This can be interpreted with a type-II superconductor analogy for the confinement in QCD, see figure \ref{superconductor}, with repulsion of the fundamental strings and with the string tension of the first topological excitation of the string (the adjoint string) larger than the double of the fundamental string tension.

The general shape of the flux-tubes observed here reinforce the conclusions of Bicudo et al. \cite{Bicudo2008a} and Cardoso et al. \cite{Cardoso2007}.

We study the fields with two geometries, U and L shaped for the Wilson loops. Here, the components of the chromoelectric and chromomagnetic fields around a static gluon-quark-antiquark are investigated. This allows us to determine the shape of the flux-tubes with respect both to the energy and action densities. When the quark and the antiquark are superposed the results are consistent with the degenerate case of the two gluon glueball, and when the gluon and the antiquark are superposed the results are consistent with the static quark-antiquark case.
Another interesting result is that the absolute value of the chromoelectric field dominates over the absolute value of the chromomagnetic field.

\acknowledgments
This work was financed by the FCT contracts POCI/FP/81933/2007 and CERN/FP/83582/2008.
We thank Orlando Oliveira for useful discussions.

\end{document}